\documentstyle{article}[12pt]

\def\O{\Omega}      
\def\o{\omega}

\def\a{\alpha}      
\def\b{\beta}     
\def\d{\delta}              
\def\D{\Delta}               
     
\def\g{\gamma}               
\def\G{\Gamma}      
\def\e{\epsilon}

\def\f{\phi}
\def\F{\Phi}
      
\def\k{\kappa}     
\def\l{\lambda}             
\def\L{\Lambda}      
\def\m{\mu}     
\def\n{\nu}      
\def\s{\sigma}      
\def\o{\omega}

\def\t{\tau}
\def\th{\theta}

\def\pa{\partial}

\def\z{\zeta}

\def\cf{{\cal F}}

\def\co{{\cal O}}

\def\car{{\cal R}}

\def\cv{{\cal V}}

\newcommand{\ti}[1]{\tilde{#1}}     
     
\newcommand{\sm}[1]{\mbox{\scriptsize #1}}      
     
\renewcommand{\@}[1]{\sqrt{#1}}     

\renewcommand{\le}[1]{\label{#1}\end{eqnarray}}      
\newcommand{\be}{\begin{equation}}
\newcommand{\ee}{\end{equation}}
\newcommand{\bea}{\begin{eqnarray}}
\newcommand{\eea}{\end{eqnarray}}
\newcommand{\nn}{\nonumber}

\newcommand{\eq}[1]{(\ref{#1})}      
\def\nn{\nonumber\\}

\def\ffract#1#2{\raise .35 em\hbox{$\scriptstyle#1$}\kern-.25em/     
\kern-.2em\lower .22 em \hbox{$\scriptstyle#2$}}

\def\half{{1\over2}\,}
\def\nonu{\nonumber \\{}}
\def\da{\dot{a}}     
\def\db{\dot{b}}

\def\tg{\tilde{\g}}

\begin{document}
\rm\large
\null\vskip-24pt
\begin{flushright}
UCLA/02/TEP/17\\
SWAT/342 \\
ITFA-2002-38\\
PUPT-2043 \\
{\tt hep-th/0210080}
\end{flushright}
\vskip0.1truecm
\begin{center}
\vskip .5truecm
{\Large\bf
On a supersymmetric completion of the $R^4$ term in \\
IIB supergravity
}
\vskip 1truecm
{\large\bf Sebastian de Haro\footnote{
e-mail: {\tt sebas@physics.ucla.edu}},
{\large\bf Annamaria Sinkovics\footnote{e-mail: {\tt sinkovic@science.uva.nl}}}
and Kostas Skenderis\footnote{
e-mail: {\tt kostas@feynman.princeton.edu}}}\\
\vskip .5truecm
${}^{1}$ {\it Department of Physics and Astronomy,   
University of California, \\
Los Angeles, Los Angeles, CA 90095, USA}
\vskip 2truemm
${}^{2}$ {\it Department of Physics,
University of Wales Swansea\\
Singleton Park, Swansea, SA2 8PP, UK} \\
\vskip 2truemm
${}^{2}\  {}^3$ {\it Institute for Theoretical Physics,
University of Amsterdam\\
Valckenierstraat 65, 1018XE Amsterdam, The Netherlands} \\
\vskip 2truemm
${}^{3}$ {\it Physics Department,   
Princeton University \\
Princeton NJ 08544, USA}
\end{center}
\vskip .3truecm
\begin{center}
{\bf \large Abstract}
\end{center}
We analyze the possibility of constructing a supersymmetric invariant 
that contains the $R^4$ term among its components
as a superpotential term in type IIB on-shell superspace. 
We consider a scalar superpotential, i.e. an arbitrary 
holomorphic function of a chiral scalar superfield.
In general, IIB superspace does not allow for the existence 
of chiral superfields, but the obstruction vanishes for a specific
superfield, the dilaton superfield. 
This superfield contains all fields of type IIB supergravity among its   
components, and its existence is implied by the solution of the Bianchi 
identities. The construction 
requires the existence of an appropriate
chiral measure, and we find an obstruction to the 
existence of such a measure. The obstruction is closely related 
to the obstruction for the existence of chiral superfields and  
is non-linear in the fields. 
These results imply that the IIB superinvariant related to the
$R^4$ term is not associated with a scalar chiral superpotential.
 
\newpage 

\tableofcontents

\newpage
\section{Introduction and summary of results}
\setcounter{equation}{0}

The effective description  of massless modes of string theories  at low energy
is given to leading order by supergravity theories. Worldsheet and string   
loops introduce higher derivative corrections to the leading order   
supergravity theory.
These terms represent the leading quantum effects of string theory
and as such are of particular importance. In particular, one may   
test duality symmetries beyond the leading order by considering
these terms. For example in the AdS/CFT duality,   
the leading higher derivative corrections of IIB string theory
are related to subleading terms in the $1/N$ and the 't Hooft coupling   
expansion in the boundary theory. Furthermore, it is of interest to   
compute stringy corrections to supergravity solutions. For instance,
one can compute stringy corrections to black hole solutions   
and their properties, such as their mass and their entropy.
This would be the leading quantum gravity effects to the   
semi-classical results. Another motivation for studying higher
derivative corrections is that they may allow to circumvent   
no-go theorems about de Sitter compactifications. Furthermore,
they may lead to stabilization of moduli in compactifications that   
to leading order yield no-scale supergravities. All of these   
applications require a detailed knowledge of the leading   
higher derivative corrections.

The higher derivative corrections can be computed systematically   
by either computing scattering amplitudes   
\cite{GW}-\cite{Peeters:2001ub} or by
using sigma model techniques  
\cite{GVZ}-\cite{AT1}.  
Despite considerable work,   
however, the complete set of the leading higher derivative terms   
is still missing. One way to extend the results in the literature
is to use the symmetries of the theory under consideration.
One such symmetry is supersymmetry. Starting from a given
term in the leading order quantum corrections
one may consider its orbit under supersymmetry.
This procedure, although straightforward, is rather challenging
from the technical point of view, and it has been carried out   
only in a few cases. Explicit results have been obtained   
for the heterotic effective action in \cite{Romans:1985xd,
Bergshoeff:1989de,deRoo:1992zp,Sch} in a component formalism and in  
\cite{Nilsson:1986rh,Gates:1986is,Gates:1986tj} in superspace.
The extension of these results to 
the other string theories as well as M theory has been discussed   
in the literature, see \cite{Peeters:2001qj} and references therein.
A complete supersymmetry analysis, however,
as well as the complete set of terms that appear at leading
order is still lacking.   

One of the most elegant ways to construct superinvariants
is to use superspace techniques.
In this  approach one is aiming in providing a superspace
formula for the higher derivative corrections.   
Upon evaluating the fermionic integration, the result   
should contain the terms obtained by sigma model and/or
scattering amplitude computations. This approach automatically   
provides all terms that are related to each other by supersymmetry.

The leading higher derivative corrections in IIB string
theory are eight derivative terms and contain the   
well-known $R^4$ term \cite{GW}.    
Dimensional analysis suggests that a superinvariant associated   
with eight derivative terms may be constructed by integrating   
a superpotential over half of superspace. Type IIB superspace
was constructed by Howe and West in \cite{HW}.
The Bianchi identities imply the equations   
of motion, so this is an on-shell superspace.
This is not a disadvantage, however.  The freedom to do field redefinitions
implies that the higher derivative terms are ambiguous
up to lower order field equations, and  by using on-shell
fields we precisely mod out by this ambiguity.\footnote{
On-shell (and off-shell) superinvariants were also extensively discussed   
in the supergravity literature in the context of possible counterterms,
see \cite{Deser:nt}-\cite{Moura:2001xx} for an (incomplete) list 
of references.}   

We should mention here that another manifestly supersymmetric
way to study the higher derivative terms is to relax some
of the supergravity constraints when solving the Bianchi   
identities. As mentioned above, the solutions of the   
Bianchi's in IIB supergravity imply the field equations.   
The equations may admit more general solutions when some
of the conventional constraints are relaxed, and one might
hope that the relaxed constraints imply the $\a'$-corrected
field equations. Such an approach for the case of M-theory
is followed in \cite{Cederwall:2000ye,Cederwall:2000fj}, see
also \cite{Cederwall:2001xk} for a review and further references.   

In \cite{HW} Howe and West   
presented a linearized superfield that satisfies a ``chirality''   
constraint (see section \ref{dilsec} for a discussion of our terminology),   
has as its leading component the physical (complex) scalar of the 
IIB supergravity,
and contains in its components all fields of type IIB supergravity.
Chiral superfields do not exist in general curved superspaces.
We show, however, that type IIB superspace allows for   
a non-linear version of the linearized superfield of Howe
and West. We call this superfield the dilaton superfield $V$.
Utilizing the solution of the Bianchi identities
given in \cite{HW}, we present iterative formulas for   
all its components.   

We then investigate the 
construction of a superinvariant as an integral over half of superspace 
of a arbitrary {\em scalar} function of the dilaton superfield. 
This choice is motivated 
by previous work of Green and collaborators, see \cite{Green:1999qt}
and references therein, where a similar construction in terms 
of the linearized superfield of Howe and West was advocated.
This is also the simplest choice for a would-be superinvariant.
Integration in superspace is a non-trivial operation:
in supergravity theories the metric transforms, so one needs to obtain an   
appropriate  supersymmetric version of the measure.
Moreover in our case the measure should be appropriate for
chiral superfields. We are thus looking for a superfield $\D$
whose leading component is $e$, where $e$ is the determinant of the 
vielbein, and is such that   
when we integrate  over half of superspace
an {\it arbitrary} holomorphic function $W[V]$ of the dilaton superfield, 
the resulting    action is supersymmetric. 

The existence of the measure is analyzed by studying the constraints
imposed by supersymmetry. We show that one can systematically   
study the cancellation of $D^n W|$ 
in the supersymmetry variation of the action starting from the terms with the
highest $n$ and moving to lower orders.  ($D$ is a supercovariant
derivatives and $A|$ denotes evaluation of $A$ at $\th=0$.)
Since the supersymmetry parameter $\z$ is complex, each $n$
leads to two different conditions. We show that 
the cancellation of the terms proportional ro $\z$   
uniquely determines all components of the measure $\D$.  
The terms proportional to $\z^*$ should  cancel automatically.
These are non-trivial conditions, and it turns out
that the cancellation does occur at leading order and at the 
linearized level at next-to-leading order, but there is 
an obstruction to the existence of the chiral measure at the 
non-linear level at next-to-leading order\footnote{
In the original version of this paper we only checked
the cancellation of the leading and linearized next-to-leading 
order terms. N. Berkovits and P. Howe subsequently found that 
certain non-linear terms at next-to-leading order 
do not cancel \cite{BH}. We thank them 
for communicating these results to us.}. The obstruction is 
very closely related to the obstruction to the existence of chiral 
superfields. This proves that there is no superinvariant 
that that can be expressed as a {\em scalar} superpotential 
of the dilaton superfield.

It turns out that the superpotential term constructed 
using the measure $\D$ determined by the cancellation of the 
$\z$ terms has rather intriguing properties, even though it is not
supersymmetric.  In the remainder we discuss these  
properties.

IIB supergravity is invariant   
under $SL(2,R)$. In string perturbation theory the $SL(2,R)$   
symmetry is broken by the vev of the dilaton, but it is   
believed that quantum theory possesses a local $SL(2,Z)$ symmetry.
It is thus of interest to analyze the $SL(2,R)$ transformation
properties of the superpotential. To this end, we   
show that the chirality condition commutes with $SL(2,R)$,
so the superpotential is compatible with $SL(2,R)$.

Given that we know all components of the dilaton superfield,   
it is straightforward but tedious to obtain the component   
form of the superpotential.  We discuss in detail
a few selected terms. In particular, we show that the   
action (in the Einstein frame) contains the terms (schematically)
\be \label{s3comp}
S_3 = \a'^3 \int d^{10} x\, e \left( t^{(12,-12)}(\t,\t^*) \l^{16}   
+ t^{(11,-11)}(\t,\t^*) \psi^* \l^{15} + \cdots   
+ t^{(0,0)} (\t,\t^*) \car^4 + \cdots \right),
\ee
where $\tau$ is the dilaton-axion,
$\l$ is the dilatino and $\psi$ is the gravitino.   
The  $\car^4$ term is given in \eq{car} and it contains
the well-known $R^4$ term, i.e.
the contraction of the indices is exactly that of the $R^4$
term that arises in string theory. The coefficients   
$t^{(12,-12)}(\t,\t^*)$, $t^{(11,-11)}(\t,\t^*)$
and $t^{(0,0)} (\t,\t^*)$ are functions of the superpotential
$W$ and its derivatives. We show by direct computation
that each $t^{(w,-w)}$ is an  eigenfunction of the $SL(2,R)$   
Laplacian acting on $(w,-w)$ forms\footnote{A $(p,q)$ \label{modular}
form $t^{(p,q)}$ transforms as   
$t^{(p,q)}(\t,\tau^*) \to t^{(p,q)}(\t,\t^*) (\g \t + \d)^p (\g \t^* + \d)^q$  
under
$ \tau \to (\a \t + \b)/(\g \t + \d)$.}. Moreover, they are related to each   
other by the application of modular covariant derivatives.
The relations between the modular forms can be  
understood as simple relations arising from the fact that  
they are derived from the same superpotential.

The eigenvalue of $t^{(0,0)} (\t,\t^*)$ under the action of the
$SL(2,R)$ Laplacian turns out to be equal to   
$20$. This eigenvalue was computed both directly, i.e.  
by acting on $t^{(0,0)} (\t,\t^*)$ with the $SL(2,R)$ Laplacian, and  
also indirectly by its relation to the eigenvalues of  
$t^{(12,-12)}(\t,\t^*)$ and $t^{(11,-11)}(\t,\t^*)$.  
Imposing $SL(2,Z)$ symmetry uniquely fixes   
$t^{(0,0)} (\t,\t^*)$ to be the Eisenstein non-holomorphic
modular function $E_5(\t,\t^*)$.  The asymptotics of   
$E_5(\t,\t^*)$ as $\tau_2 \to \infty$ is   
$\tau_2^5$ and $\tau_2^{-4}$ plus exponentially suppressed   
terms\footnote{The powers in the asymptotic terms   
are directly linked to the eigenvalue of Laplacian. A  
short computation shows that $\nabla_{SL(2)} \tau_2^s = s(s-1) \tau_2^s$,
so with eigenvalue $20$, one has that each of $\tau_2^5$ and   
$\tau_2^{-4}$ are eigenfunctions.}.   
This implies that the string frame
effective action at ``weak coupling'' contains the terms,
\be  \label{s3fin}
S = \int d^{10} x \sqrt{g} [e^{-2 \varphi} R + ... +   
 \a'^3 (c_1 e^{-{11 \over 2} \varphi}  + c_2 e^{{7 \over 2} \varphi})\,  
{\cal R}^4 + ...]~,
\ee
where $c_1$ and $c_2$ are non-zero numerical constants.
This asymptotic behavior is not consistent with  
IIB string theory (as we know it).
In closed string perturbation theory the leading behavior is   
$g_s^{-2}$ and only even powers of $g_s$ appear. The leading behavior
in our case is more singular than the string tree-level contribution
and is half-integral. Moreover, the   
difference of the two ``perturbative'' contributions is an   
odd power of $g_s$, i.e. $g_s^9$, so even if one would normalize
by hand the leading power to be one (which $SL(2,Z)$ does not allow),
the resulting series would still be inconsistent with closed
string perturbation theory. Open string loops can give
odd powers of $g_s$ but in our computation there are no open strings.

One should contrast these results to other results reported
in the literature. In \cite{Green}
Green and Gutperle  conjectured that the coefficient
of the $R^4$ term that arises in string theory 
is the non-holomorphic Eisenstein series $E_{3/2}$.
The asymptotics of this modular function is consistent with tree-level
and one-loop contributions and implies that there are no further
perturbative contributions. Supporting evidence for this conjecture
as well as related works appeared in
\cite{Green:1997me}-\cite{Chalmers:2000zg},
see \cite{Green:1999qt} for a review.   
In particular, Berkovits constructed in \cite{Berkovits:1997pj} an invariant   
that contains the $R^4$ term in $N=2$ $d=8$ {\it linearized}   
superspace. This theory can be viewed as a $T^2$ reduction of   
type IIB supergravity. He also showed that it contains exactly   
tree-level and one-loop contributions. Pioline in \cite{Pioline:1998mn}
showed that the superinvariant constructed in the $8d$ linearized superspace
is an eigenmode of the Laplacian with eigenvalue that implies
that the coefficient of the $R^4$ term is $E_{3/2}$.
Finally, Green and Sethi \cite{GrSe} analyzed the supersymmetry  
constraints on the coefficients of $\l^{16}$ and $\psi^* \l^{15}$ terms.
Their analysis involved specific $\a'$ corrections to the supersymmetry
rules and with these corrections they showed  
that the coefficient of the $R^4$ term is the Eisenstein
series $E_{3/2}$.

As discussed, the superpotential term in our case is not supersymmetric, 
so our results are not in conflict with existing results.
Our results, however, imply that the tree- and one-loop $R^4$ terms
are not associated with a scalar superpotential term,
even at the linearized level.
Notice that our results do not rule out that the superpotential
term is invariant under {\em  linearized} supersymmetry --
the obstruction is non-linear in the fields. It is the 
fact that we get $E_5$ rather than $E_{3/2}$ as a coefficient 
of the $R^4$ term that implies that the superpotential term 
is not associated with the stringy $R^4$ term.
Notice that these considerations are consistent with   
the discussion in \cite{Peeters:2001ub} where it was argued
that certain terms obtained by string amplitude computations cannot   
be part of the superinvariant based on the linearized   
superfield of Howe and West.   

In this paper we investigate whether 
the superinvariant associated with the $R^4$ term
can be constructed as an integral over half of superspace
of a {\em scalar} superpotential. Even though the answer
turned out to be negative, we believe that the 
techniques developed here will be useful for the construction
of the actual superinvariant associated with the $R^4$ term.
Our analysis was specific to IIB supergravity.
The method, however, is general and we expect that similar
constructions apply to the other string theories and to M-theory.   
The superspace for type IIA theory has been   
constructed in \cite{Carr:1986tk}, of type I supergravity   
coupled to Yang-Mills in   \cite{Nilsson:1981bn}-\cite{Tak},   
and of eleven dimensional supergravity   
in \cite{Cremmer:1980ru}-\cite{Howe:1997rf}. A construction
similar to ours for the superinvariant associated with the $R^4$
terms in type I theories has been discussed in   
\cite{Nilsson:1986rh}. Although the details in all these   
cases will be different, we expect that   
one has to go through the same steps we present here.   
Keeping this in mind, we shall present in some detail
the development of the tools required in order to carry   
out the computations.
 
This paper is organized as follows. In the next section we   
discuss the set up of the computation. In particular, we discuss   
the issue of working with on-shell fields. In section \ref{review}
we review type IIB supergravity both in components and in superspace.
Chiral superfields and in particular the construction of the   
dilaton superfield and its components are discussed in section
\ref{dilsec}. The (non-existence of the) chiral measure is   
discussed in section \ref{saction}. Section \ref{sl2} presents the   
computation of the components of the superpotential as well   
as the $SL(2,R)$ transformation properties. We end with the   
discussion of our results and of future directions.   

An effort is made to make this paper self-contained.
Several appendices summarize relevant results from  
the literature. In appendix A we discuss   
our conventions, in appendix B we summarize the solution   
of the Bianchi identities, and in  appendix C we provide
the supersymmetry rules of the type IIB supergravity.  
Appendices D and E present details of computations  
used in the main text. In particular, in   
appendix D we discuss the $F_5$ dependence of the dilaton
superfield and in appendix E we show that the
superpotential contains the well-known $R^4$ term.

\section{Symmetries of the effective action} \label{sym}
\setcounter{equation}{0}

We discuss in this section the constraints imposed on low energy
effective actions by symmetries. Recall that the   
low energy effective actions are expressed as an expansion in
derivatives (with the dimension of various fields   
properly taken into account). In string theories   
the various terms are weighted by different powers of $\a'$.
The effective action has thus the form
\be
S=S_0 + \sum_{n \geq 3} \a'^n S_n~,
\ee
where we take the sum to start at $n=3$ because in the type IIB   
string theory this is the leading correction.   

Let us consider now a symmetry $\d_0$ of $S_0$. The symmetry   
transformation may also receive corrections,
\be
\d = \d_0 + \sum_{n \geq 3} \a'^n \d_n~.
\ee
Invariance of the effective action, $\d S = 0$, then implies    
\bea
\d_0 S_0 &=& 0 \nn
\d_0 S_3 + \d_3 S_0 &=& 0~, \label{s3inv}   
\eea
etc. This procedure constrains both the possible terms
$S_3$  and also the possible deformations of the symmetry
$\d_3$.

The higher derivative terms, however, are ambiguous due to   
field redefinitions \cite{Tseytlin:1986zz,GW,Gross:1986mw}.
Let us call collectively $\f^I$   
all the fields. Then the field redefinition,
\be \label{redef}
\f^I \to \f^I + \a'^3 f^I(\f^J)
\ee
where $f^I(\f^J)$ an an arbitrary non-singular function of $\f^I$,
implies that   
\be
S_3 \to S_3 + f^I {\pa S_0 \over \pa \f^I}~.
\ee
Thus, $S_3$ is ambiguous up to lowest order field equations.
To eliminate this ambiguity one may work with on-shell
fields.

We now show that instead of solving (\ref{s3inv}), one may   
equivalently solve   
\be \label{s3on}
\d_0 S_3 \approx 0~,
\ee
where $\approx$ means that the equality is up to the lowest   
order field equations (i.e. the field equations that follow
from $S_0$). Indeed, any solution of (\ref{s3inv}) is also   
a solution of (\ref{s3on}). To see this we use the chain   
rule to rewrite (\ref{s3inv}) as
\be   
\d_0 S_3 = - {\d S_0 \over \d \f^I} \d_3 \f^I \quad
\Rightarrow \quad \d_0 S_3 \approx 0~.
\ee
Conversely, for any solution of (\ref{s3on}) there exists some
$\d_3$ such that (\ref{s3inv}) is satisfied. To see this we   
note that (\ref{s3on}) means that   
\be \label{s3on2}
\d_0 S_3= g^I {\d S_0 \over \d \f^I}
\ee
for some function $g^I$ of all fields.
Now let us consider the following deformation of $\d_0$
\be \label{d3}
\d_3 \f^I = - g^I~.
\ee
Then (\ref{s3on2}) implies that (\ref{s3on}) is satisfied.
We thus established that   
\be
(\d_0 + \a'^3 \d_3) (S_0 + \a'^3 S_3) = \co(\a'^4)
\ee
with $\d_3$ given by (\ref{d3}).
This then implies
\be
[\d, \d] S = \co(\a'^4)~.
\ee
Thus $[\d, \d]$ is a symmetry of the action and the
algebra of $\d$ closes up to this order (in the absence of  
auxiliary fields the algebra  
will only close on-shell). This finishes the proof that one   
may consider either (\ref{s3inv}) or (\ref{s3on}).

The above discussion also gives a prescription for calculating   
explicitly the deformation of the symmetry.   
One first solves the problem (\ref{s3on})   
in terms of on-shell fields. After $S_3$   
is determined, one computes $\d_0 S_3$   
with the on-shell condition relaxed,   
and finally reads off $\d_3 \f^I$ from (\ref{s3on2})   
and (\ref{d3}).   

Notice also that there may be more than one   
(inequivalent) solutions of the problem (\ref{s3inv})
(or (\ref{s3on})). In other words, there may be   
different pairs $\d_3, S_3$ that satisfy   
(\ref{s3inv}).

Because of the freedom of field redefinitions
not all $\d_3 \f^I$ are non-trivial. Indeed some   
of them may be removed by a field redefinition.
By this we mean that after the field redefinition (\ref{redef})
the action will be supersymmetric (i.e. (\ref{s3inv}) will hold) without the  
need for $\d_3 \f^I$.  
A condition for this to happen is that the   
variation is of the form
\bea \label{trivial}
\d_3 (\e) \f^I &=& {\cal L}_{f^I} (\d_0(\e) \f^I) + \d_0(\e') \f^I \nn
&=&- f^J \pa_J (\d_0 (\e) \f^I) + \pa_J f^I (\d_0(\e) \f^J)
                 + \d_0(\e') \f^I~,
\eea
where $\e$ is the parameter of the variation, $\e'$ is a possibly field
dependent parameter, ${\cal L}_{f^I}$ is the Lie derivative in the   
space of fields, and $\pa_I = \d/\d \f^I$.  

\section{Review of IIB supergravity} \label{review}
\setcounter{equation}{0}

\subsection{Component formulation}

Type IIB supergravity was constructed in \cite{S,HW}.
The bosonic field content consists of the   
metric $g_{m n}$, a complex antisymmetric two-form   
gauge field $a_{m n}$, a real four-form gauge field   
$a_{nrst}$ with a self-dual field strength (at the linearized level),   
and a complex scalar $a$. The fermions consist of a complex
gravitino $\psi_m$ of negative chirality and a complex dilatino $\l$
of positive chirality (our conventions are given in appendix \ref{conv})
\be
\G_{11} \psi_m = - \psi_m, \qquad \G_{11} \l = \l~.
\ee

The covariant field equations are invariant   
under an $SU(1,1)$ global symmetry which is realized   
non-linearly on the scalars. To realize the $SU(1,1)$
linearly we add an auxiliary scalar and
an extra $U(1)$ gauge invariance \cite{SW}. The extra   
scalar may be eliminated by fixing the $U(1)$ gauge   
invariance. We shall work with the gauge invariant   
formulation throughout this paper.

The scalars parametrize the coset space $SU(1,1)/U(1)$.   
We represent them as an $SU(1,1)$ group matrix,
\be \label{nu}
\cv=
\left(
\begin{array}{cc}
u & v \\
v^* & u^*
\end{array}
\right)
\ee
where $u u^* - v v^*=1$.
The global $SU(1,1)$ acts by a left multiplication, and the local   
$U(1)$ by matrix multiplication from the right,
\be \label{su11}
\cv'=   
\left(
\begin{array}{cc}
z & w \\
w^* & z^*
\end{array}
\right)
\left(
\begin{array}{cc}
u & v \\
v^* & u^*
\end{array}
\right)
\left(
\begin{array}{cc}
e^{-i \Sigma} & 0 \\
0 & e^{i \Sigma}
\end{array}
\right)~.
\ee
The components $v$ and $u^*$ have $U(1)$ charge
$U=1$ and $u$ and $v^*$ charge $U=-1$.
It follows that the (complex) ratio   
\be
a={v \over u^*}
\ee
is gauge invariant and   
represents the physical scalars of the theory. Under $SU(1,1)$ it transforms   
with a linear fractional transformation,
\be   
a' = {z a  + w \over w^* a + z^*}~.
\ee
The complex scalar $a$ parametrizes the unit disc.
The axion and dilaton of string theory are related to $a$ by
a (non-linear) transformation that we describe below.

The metric and the four-form are inert under $SU(1,1)$   
and neutral under $U(1)$. The   
antisymmetric tensor is neutral under $U(1)$, and
transforms as a doublet under
$SU(1,1)$, where the two components of the   
corresponding column vector are $a^*_{m n}$ and   
$a_{mn}$. The gravitino has charge $U=1/2$ and the dilatino
has $U=3/2$. Both of them are inert under   
$SU(1,1)$. Finally the supersymmetry parameter $\z$   
is neutral with respect to $SU(1,1)$ and has $U=1/2$.

It will also be useful to discuss the transformation properties   
of the fields under the $U(1)$ subgroup of $SU(1,1)$. These transformations
can be obtained from the $SU(1,1)$ transformations by setting
$z=\exp i \s, w=0$. In particular, $u$ and $v$ have charge 1 and   
$u^*$ and $v^*$ have charge -1. It follows that the physical   
scalar $a$ has charge 2. Notice that $a$ is invariant under the local   
$U(1)$ symmetry but transforms under the $U(1)$ subgroup of $SU(1,1)$.
The dilatino and gravitino have charge 3/2 and 1/2, respectively.
We summarize the results about the two $U(1)$ charges in the following table:
\begin{table}[h]
\begin{center}
\begin{tabular}{|c|c|c|c|c|c|c|c|c|c|c|c|c|}
\hline
\ & $g_{mn}$ & $\psi_m$ & $\l$ & $a_{mn}$ & $a_{mnrs}$ & $u$ & $v$ &   
$u^*$ & $v^*$ & $a$ & $\z$ \\   
\hline
local $U(1)$  & 0 & 1/2 & 3/2 & 0 & 0& -1 & 1 & 1 & -1 & 0 & 1/2 \\
\hline
global $U(1)$  & 0 & 0 & 0 & 1 & 0& 1 & 1 & -1 & -1 & 2 & 0 \\
\hline      
\end{tabular}
\caption{{\it The charges of the fields under the local $U(1)$ and
the $U(1)$ subgroup of $SU(1,1)$.}
}
\end{center}
\end{table}

The field equations are constructed using $SU(1,1)$ invariant
combinations of the scalars and the 2-form $a_{mn}$.
The 1-forms
\be \label{PQ}
p=u^* d v - v d u^*, \qquad q={1 \over 2 i} (u^* d u - v d v^*)
\ee
are $SU(1,1)$ invariant.
The composite field $q$ transforms as a connection under the   
local $U(1)$, and $p$ has charge 2.
Let   
\be
\cf = da_2
\ee
be the field strength associated with $a_{mn}$.
The $SU(1,1)$ invariant field strengths can be constructed as
\be
(f^*_{3}, f_3) = (\cf^*, \cf) \cv
\ee
We finally introduce
\be
f_5 = d a_4 - 2i (a_2^* \wedge \cf - a_2 \wedge \cf^*)
\ee

As mentioned above, we will work with the gauge invariant
formulation throughout (except in section \ref{sl2}).   
We briefly discuss here gauge fixing.
More details can be found in \cite{S} (where a different
realization is used). An explicit realization of $\cv$ is given by
\be \label{real}
\cv={1 \over \sqrt{1 - a a^*}}   
\left(
\begin{array}{cc}
e^{-i \phi} & a e^{i \phi} \\
a^* e^{-i \phi} & e^{i \phi}
\end{array}
\right)~.
\ee
The local $U(1)$ symmetry leaves invariant $a$ and acts by shifting
$\phi$. It follows that one may gauge fix the $U(1)$ symmetry   
by setting $\phi$ equal to some function of $a$,
see (\ref{gfix}) for the gauge fixing we will use later.
$SU(1,1)$ and supersymmetry transformations do not respect this gauge, so   
compensating $U(1)$ transformations are necessary. In particular, this   
implies that the fermions that were $SU(1,1)$ invariant in the gauge invariant
formulation, transform after gauge fixing.  The compensating  
transformations complicate the computations. We therefore choose to  
work throughout in the gauge invariant formulation. Only in section  
\ref{sl2}, where we will compare our results with results  
in the literature, we will gauge fix the $U(1)$.  
  
In our discussion so far we took the scalars to parametrize
the $SU(1,1)/U(1)$ coset space. From the string theory point   
of view, it is more appropriate to consider the (equivalent)
description where the scalars $\tau$ parametrize the $SL(2,R)/SO(2)$
coset. To go from one description to another we note that the   
$a$ parametrizes the unit disc, whereas $\t$ the   
Poincare upper half plane. The transformation from one to another is   
given by
\be \label{tau}
\tau=i{1-a \over 1+a}~.
\ee
The fact that $\t$ rather than $a$   
is related to the axion $C_0$ and dilaton $\varphi$   
of type IIB string theory by   
$\t =\t_1 + i \t_2 = C_0 + i \exp(-\varphi)$
is explained, for instance, in \cite{BHT}.
The $SU(1,1)$ transformation given in (\ref{su11})
is related to an $SL(2,R)$ transformation by   
\bea \label{sl2b}
&&\a=z_1-w_1,\,\,\, \qquad \b=z_2+w_2, \nn
&& \g=-z_2+w_2, \qquad \d=z_1+w_1,
\eea
where $z=z_1+i z_2, w=w_1+i w_2$. Using these  
results one can go from one formulation to another.
For example, $\t$ transforms as
\be
\t' = {\a \t + \b \over \g \t + \d}~.
\ee       
Notice also that $\t$ does not transform linearly under the   
$SO(2)$ subgroup of $SL(2,R)$ (the $SO(2)$ subgroup is generated by
$\a=\d=\cos \s, \ \g=-\b=\sin \s$). The infinitesimal transformation
is given by   
\be
\delta \t=\s (1-\t^2)~.
\ee
In contrast $a$ transforms
linearly under the same $SO(2)$ (it has charge 2, as we explained).

\subsection{Superspace formulation}   

\subsubsection{General set-up}

IIB superspace was constructed in \cite{HW}.
Our superspace conventions are the ones of \cite{HW} and   
are summarized in appendix \ref{conv}. We denote the   
superspace coordinates by $z^M = (x^m, \th^\mu, \th^{\bar{\mu}})$,
where $x^m$ are the spacetime coordinates and
$\theta^{\mu}$ is a complex 16-component Weyl spinor   
of $SO(9, 1)$ and $\th^{\bar{\mu}}$ is its complex conjugate.
For every $x$-space field we introduce a superfield
whose leading component is the $x$-space field.
We will denote the superfield with the same letter as the $x$-field but   
capitalized.   

The $SU(1,1)$ group matrix $\cv$ becomes a superfield, and
the scalar fields $u$ and $v$ are   
the lowest components of the superfields $U$ and $V$, respectively.
They satisfy   
\be \label{det}
U U^* - V V^*=1~.
\ee
We also define the superspace version of (\ref{PQ}),
\be \label{PQS}
P=U^* d V - V d U^*, \qquad Q={1 \over 2 i} (U^* d U - V d V^*),
\ee
where we use a form notation.
It will be useful to give these in components as well. For the composite
connection we get
\be
Q_A = {1 \over 2 i} (U^* \pa_A U - V \pa_A V^*) =   
- {1 \over 2 i} (U \pa_A U^* - V^* \pa_A V)~,
\ee
where in the second equality we used (\ref{det}).
Using the definition of the covariant derivative,
\be
D_A V = \pa_A V + 2 i Q_A V, \qquad   
D_A U = \pa_A U - 2 i Q_A U,
\ee   
we derive the identities,
\be \label{idUV}
U^* D_A U = V D_A V^*, \qquad U D_A U^* = V^* D_A V~.
\ee
Using these identities we then obtain
\bea
&&P_A = U^* D_A V - V D_A U^* = {1 \over U} D_A V, \nn
&&\bar{P}_A = U D_A V^* - V^* D_A U = {1 \over U^*} D_A V^*~.
\label{pdef}
\eea

The theory is invariant under local $SO(1,9) \times U(1)$
transformations that rotate the frame fields
$E^A = d z^M E_M^A$. The $SO(1,9)$ is the $10d$ Lorentz
group and the $U(1)$ is identified with the local $U(1)$   
of $SU(1,1)/U(1)$. There is a corresponding 1-form
connection $\O_A{}^B$. The $SO(1,9)$ part is a superfield
with lowest component (a supercovariant generalization   
of) the $x$-space spin-connection (see (\ref{spincon}))   
and the $U(1)$ part is a superfield $Q_A$.

The superspace geometry is encoded in the algebra of   
supercovariant derivatives,
\be
[D_A, D_B \} = - T_{A B}{}^C D_C   
+ \frac{1}{2} R_{ABC}{}^D L_D{}^C + 2 i M_{AB} \k
\ee   
where $T_{AB}{}^C$ is the torsion, $L_A{}^B$ are   
the $SO(9,1)$ generators, $\k$ is the $U(1)$ generator  
 and $R_{ABC}{}^D$ and $M_{AB}$  are the spacetime   
and $U(1)$ curvature tensors, respectively.
The super-Jacobi identity
\be
[D_A, [D_B, D_C\} \} + {\rm graded \ cyclic} =0
\ee
implies the Bianchi identities  
\bea
I^{(1)}_{ABC}{}^{D} &=& \sum_{(ABC)} (D_A T_{BC}{}^{D} + T_{AB}{}^{E}   
T_{EC}{}^D - \hat{R}_{ABC}{}^{D}) = 0 \label{b1} \\
I^{(2)}_{ABCD}{}^{E} &=& \sum_{(ABC)} (D_A \hat{R}_{BCD}{}^{E} + T_{AB}{}^F   
\hat{R}_{FCD}{}^E ) =0 \label{b2}
\eea
where $\sum_{(ABC)}$ denotes the graded cyclic sum,
and the hat in $\hat{R}$ means that it contains the contribution
of the $U(1)$ connection as well.
There are also additional identities that stem from the   
fact that field strengths are closed forms.
In form notation \cite{HW},  
\bea
I^{(3)} &=& D F_3 - F_3^* \wedge P, \label{b3} \\
I^{(4)} &=& d F_5 + 2 i F_3 \wedge F_3^* \label{b4} \\
I^{(5)} &=& D P \label{b5} \\
I^{(6)} &=& M + \half i P \wedge P^* \label{b6}~.
\eea

The superfields introduced so far have a large number of   
components. In order to reduce the independent fields to the   
ones of the IIB supergravity multiplet
described in the previous section one needs to impose   
constraints. Once the constraints are imposed the   
equations (\ref{b1})-(\ref{b6}) are not identities
any more and should be solved. Solving the Bianchi   
identities determines the corresponding superspace.

The Bianchi identities for IIB supergravity  were   
solved in \cite{HW}. In superspace it is the   
torsion rather the curvature that is more important.
The curvature is determined once the torsion coefficients are   
supplied.  The non-zero torsion components are   
specified by the field content of the IIB supergravity.
For each field of IIB supergravity there   
is a torsion coefficient whose leading component   
at the linearized level is the corresponding field strength.
At the non-linear level the torsion coefficients   
contain fermion bilinears as well. The exact expressions are   
collected in appendix \ref{bianchi}. Furthermore,   
the Bianchi identities imply the IIB field   
equations. Notice that we do not need to consider $\alpha'$ corrections
to the torsion constraints, since our method makes use of the lowest
order supersymmetry transformations only. However, as discussed
in section \ref{sym}, a specific superinvariant does imply specific
corrections to the supersymmetry rules, and the latter  
induce $\a'$ corrections to the torsion coefficients.

\subsubsection{Solution of the Bianchi identities}

To completely determine the theory we need the fermionic derivatives
of all fields. This information can be obtained from 
Bianchi identities. We summarize these results here.   
The derivation of most of the formulas below can be found   
in \cite{HW}.

The identities $I^{(5)}$ and $I^{(6)}$ imply,
\be
P_{\bar{\a}} = 0, \qquad P_\a = - 2 \L_\a,  \qquad
\bar{P}_{\a}=0, \qquad \bar{P}_{\bar{\a}} = - 2 \L^*_\a~.
\ee
Using (\ref{pdef}), these results imply
\bea \label{ch}
&&D_\a^* V = 0,   \qquad D_\a V = - 2 U \L_\a \qquad
D_\a^* U^* = 0, \qquad D_\a U^* = - 2 V^* \L_\a \nn
&&D_\a V^* = 0,   \qquad D_\a^* V^* = - 2 U^* \L_\a^* \qquad
D_\a U = 0, \qquad D_\a^* U = - 2 V \L_\a^*~.   
\eea   

A direct computation starting from the definition (\ref{pdef})   
gives the fermionic derivatives of $P_a$,
\bea
D_\a P_a&=&-2D_a\L_\a -2T_{a\a}{}^\b\L_\b \nn
D_\a\bar P_a&=& -2T_{\a a}{}^{\bar\b} \L^*_\b~.
\eea
Notice that $P_a^* = - \bar P_a$.

The dimension one\footnote{
The dimension of the Bianchi identity $I_{ABC}{}^D$   
is equal to $A+B+C-D$, where a bosonic index counts as 1   
and a fermionic one as 1/2.}
 Bianchi identities $I^{(1)}_{\a \b \g}{}^\d,   
I^{(1)}_{\a \b \g}{}^{\bar{\d}},   
I^{(1)}_{\a \b c}{}^d$, and $I^{(1)}_{\a \b \g}{}^\d$,   
determine the fermionic derivatives of $\L_\a$,
\bea
D_\a\L_\b&=&-{i\over24}\,\g^{abc}_{\a\b}F_{abc} \label{DL} \\
D_{\a} \L^*_\b &=& - \half i (\g^a)_{\a \b} \bar{P}_a~. \label{bDL}
\eea

The dimension 3/2 identities   
$I^{(1)}_{a \b \g}{}^\d, I^{(1)}_{\a \b \bar{\g}}{}^{\bar{\d}},   
I^{(1)}_{a \b \g}{}^{\bar{\d}}$ yield the fermionic derivatives
of $F_3$ and $F_5$, but the fermionic derivative of   
$F_5$ is easier to obtain from $I^{(4)}$.
The results are   
\bea
D_\a F_{abc}&=&-{1\over32}\left(\g_{abcdef}\bar F^{def}+3
F^*_{[a}{}^{de}\g_{bc]de} +52 F^*_{[ab}{}^d\g_{c]d} +28
F^*_{abc}\right)_\a{}^\b \L_\b  \nn
&&+3 P_{[a}\g_{bc]\a}{}^\b\L^*_\b +3i\g_{[a\a\b}\Psi_{bc]}{}^\b, \label{dF} \\
D_\a F^*_{abc}&=&-3(\g_{[ab})_\a{}^\b D_{c]}\L_\b^* +3T_{[a\a}{}^\b
\g_{bc]\b}{}^\g \L_\g^* \label{dbF} \\
D_{\a} F_{abcde}&=&-10 \Psi_{[ab}{}^{\bar{\b}} \g_{cde]\a \bar{\b}}   
+20  (\g_{[ab} \L^*)_\a F_{cde]}~.
\eea
where $\Psi_{ab}{}^\a$ is a superfield whose leading component
is the covariantized field strength of the gravitino.

The dimension two Bianchi's $I^{(1)}_{a b \g}{}^\d$ and   
$I^{(1)}_{a b \g}{}^{\bar{\d}}$
give
\bea
D_\a \Psi_{bc}{}^{\e} &=& \frac{1}{4} (\g^{ef})_{\a}{}^{\e} R_{bcef} -
 \left( D_b T_{c \a}{}^{\e} + T_{b \a}{}^{\k} T_{c \k}{}^{\e} -     
T_{b \a}{}^{\bar{\k}} T_{c \bar{\k}}{}^{\e} - ( b \leftrightarrow c) \right)
+ i \d^{\e}_{\a} M_{bc}, \label{DFDPDPsi} \nn  
D_\a\Psi_{ab}{}^{\bar\b}&=&-2D_{[a}T_{b]\a}{}^{\bar\b} -\Psi_{ab}{}^\g
T_{\g\a}{}^{\bar\b} +2T_{\a[a}{}^\g T_{b]\g}{}^{\bar\b}
-2T_{\a[a}{}^{\bar\g} T_{b]\bar\g}{}^{\bar\b}~. \label{dPsib}
\eea

Finally, the fermionic derivative of the Riemann tensor   
can be determined from $I^{(2)}_{\a b c d}{}^e$,
\be
D_\a R_{bcd}{}^e=-2D_{[b} R_{c] \a d}{}^e   
-2T_{\a [b}{}^{\bar\g} R_{c]\bar\g d}{}^e   
+2T_{\a [b}{}^\b R_{c] \b d}{}^e   
-T_{bc}{}^\b R_{\a\b d}{}^e
+T_{bc}{}^{\bar\b}R_{\a\bar\b d}{}^e~.
\ee

\subsubsection{From superspace to components}

The components of a covariant superfield $S$ may be
obtained by the method of covariant projections, i.e.   
the components of $S$ are obtained by evaluating   
successive spinorial covariant derivatives at
$\th=0$:
\be
s \equiv S|_{\th=0}, \qquad s_\a \equiv D_\a S|_{\th=0}, \qquad
s_{\bar{\a}} \equiv D^*_{\a} S|_{\th=0},   
\qquad {\rm etc.}
\ee
For ease in notation we will denote the projection by a vertical
line without the subscript $\th=0$.   

The vielbein and the gravitino are given by
\be
E_m^a| = e_m^a \qquad
E_m^\a| = \psi_m^\a ~.   
\ee
To compute the components of a tensor whose indices have been converted
to target space indices one uses manipulations of the form
\be \label{flcu}
X_a| = E_a^M| X_M| = e_a{}^m X_m|   
- \psi_a^\a X_\a| + \psi^*_a{}^\a X_{\bar{\a}}|~.
\ee
Using such manipulations one can compute the leading   
components of the superfields $P_a, F_{abc}, F_{abcde}$.
These are the supercovariant field strengths which   
we denote by hats,
\bea \label{scov}
P_a| &=& \hat{p}_a = p_a + 2 (\psi_a \l) \nn
F_{abc}| &=& \hat{f}_{abc} = f_{abc} - 3 (\psi^*_{[a} \g_{bc]} \l)   
- 3 i (\psi_{[a} \g_b \psi_{c]}) \nn
F_{abcde}| &=& \hat{f}_{abcde} = f_{abcde} + 20   
(\psi^*_{[a} \g_{bcd} \psi_{e]})~.
\eea
Similarly, one obtains the supercovariant version of the   
spin-connection, gravitino field strength and Riemann   
tensor,   
\bea \label{spincon}
\O_{mab}|&=& \hat{\o}_{mab}= \o_{mab}(e) + \k_{mab},  \nn
\Psi_{ab}{}^{\bar{\a}}| &=& \hat{\psi}_{ab}{}^{\bar{\a}}              
=2 e_a^m e_b^n D_{[m} \psi_{n]}{}^{\bar{\a}}   
- 2 \psi_{[a}{}^\b T_{b]\b}{}^{\bar{\a}}   
+ 2 \psi^*_{[a}{}^{\bar{\b}} T_{b]\bar{\b}}{}^{\bar{\a}}
+\psi_a \g^c \psi_b (\g^c \l^*)^\a - 2   
\psi_{[a}^\a \psi_{b]}^{\bar{\b}} \l^*_\b \nn
R_{abc}{}^d|&=&\hat{r}_{abc}{}^d =   
r_{abc}{}^d - 2 \psi_{[a}{}^\a R_{b]\a c}{}^d|
+2 \psi^*_{[a}{}^{\bar{\a}} R_{b]\bar{\a} c}{}^d|   
+\psi_{a}{}^\a \psi_{b}{}^\b R_{\b \a c}{}^d| \nn
&& \qquad \quad
+\psi^*_{a}{}^{\bar{\a}} \psi^*_{b}{}^{\bar{\b}} 
R_{\bar{\b} \bar{\a} c}{}^d|   
-2 \psi^*_{[a}{}^{\bar{\a}} \psi_{b]}{}^\b R_{\b \bar{\a} c}{}^d|~,   
\eea
where $\o_{amn}(e)$ is the standard spin-connection associated with the   
vielbein, and   
\be
\k_{a,bc} = \half (t_{ab,c} + t_{ca,b} - t_{bc,a}),  
\qquad t_{mn}{}^a = - 2 i \psi^*_{[m} \g^a \psi_{n]}~.
\ee

General coordinate transformations in superspace with   
parameter $\zeta^a \equiv \xi^\a(z)|$ yield the   
local supersymmetry rules for the components fields.
For a covariant superfield $S$, this yields
\be \label{susy1}
\d S| = \z^\a (D_\a S)| - \z^\a{}^* (D^*_\a S)|~.
\ee
The supersymmetry rule for the vielbein and gravitino
is obtained using    
\be \label{susy2}
\d E_M{}^A = D_M \xi^A - E_M{}^C \xi^B T_{BC}{}^A~.
\ee
The supersymmetry rules for the component fields
are collected in appendix \ref{st}.

\section{The dilaton superfield} \label{dilsec}
\setcounter{equation}{0}

\subsection{General construction}

Chiral superfields in four dimensions satisfy the linear   
constraint, $D_{\dot{\a}} \Phi=0$. In ten dimensions one   
may  attempt to impose the linear constraint,
\be \label{chiral}
D^*_\a \Phi=0~.
\ee
A superfield  satisfying such a constraint may be   
called an ``analytic superfield'' since in the chiral
representation the superfield depends only on $\th$
but not $\th^*$. Because of the similarity with the   
$4d$ chiral superfields, however, we will still   
use the terminology ``chiral superfield'' even though   
it is not appropriate in $10d$ where   
$D^*_\a$ and $D_\a$ have the same chirality.   

In flat superspace one can always impose the constraint
(\ref{chiral}). In curved spacetime, however, there is   
an integrability condition: the anti-commutator of   
two $D^*_\a$ acting on the superfield should also vanish. If the
torsion $T_{\bar{\a} \bar{\b}}^\g$ is non-zero   
then the chirality constraint (\ref{chiral})   
cannot be in general imposed. Indeed,   
\be
0=\{ D^*_\a, D^*_\b \} \Phi = - T_{\bar{\a} \bar{\b}}^\g D_\g \Phi~.
\ee   
In IIB supergravity,
\be
T_{\bar{\a} \bar{\b}}^{\g} = (\g^{a})_{\a \b} (\g_a)^{\g \d} \L_\d -   
2 \d_{(\a}^{\g} \L_{\b)}~.
\ee
So this equation yields
\be
\g^{a}_{\a \b} \L_\g (\g_a)^{\g \d} D_\d \Phi   
- 2 \L_{(\a} D_{\b)} \Phi=0~,
\ee
which implies
\be
\L_{(\a} D_{\b)} \Phi=0~.   
\ee
The most general solution of this equation is   
\be \label{chL}
D_\a \Phi = g \L_\a~,   
\ee
where $g$ is any function of the fields. Notice also that if $\Phi$
is a chiral superfield then its covariant derivatives $D_a \Phi$ 
and $D_\a \Phi$ are not. This follows from the explicit form of the
(anti)commutator of $D_\a^*$ with $D_a$ and $D_\a$.   

We have seen in the previous section that the Bianchi identities
imply that $V$ and $U^*$ are chiral superfields (and   
$V^*$ and $U$ antichiral ones), see (\ref{ch}).
Indeed, in this case (\ref{chL}) is satisfied
with $g=-2 U$ and $g=-2 V^*$, respectively. These
two superfields are not independent. From (\ref{idUV})
we get   
\be \label{du*}
D_\a U^* = {V^* \over U} D_\a V~.
\ee
Clearly, any function
of $V$ and $U^*$ (but not of their covariant derivatives)
is a chiral superfield as well. The gauge invariant   
superfield,
\be
A={V \over U^*}
\ee
is chiral and has as its lowest component the physical scalar
fields $a$. This superfield has $U(1)$ charge 2 with respect to
the $U(1)$ subgroup of $SU(1,1)$. The linearized version of this superfield
was constructed in \cite{HW}. It was shown there that it contains
the entire supergravity multiplet in its components. Our considerations
lead to the construction of all components, including all   
non-linear terms. Another related superfield is   
\be
T = i {1 - A \over 1+A}~.
\ee
The leading component of this superfield is $\tau$.

As discussed in the previous section, one can obtain   
all components of a superfield by a successive fermionic
differentiation and then evaluation at $\th=0$.   
To obtain the components of all superfields discussed
in this paper, it is sufficient to compute the fermionic derivatives of $V$
and this is the subject of the next subsection.

Before we proceed, however, we show that the chirality 
condition commutes with the $SU(1,1)$ (or equivalently $SL(2,R)$)
action. From (\ref{su11}) we get that $SU(1,1)$ acts on the 
chiral superfields, $V$ and $U^*$ as
\be
V' = z V + w U^*, \qquad U^*{}' = w^* V + z^* U^*
\ee
We thus see that the $SU(1,1)$ transformation rotates the 
two chiral superfields among themselves, and therefore the transformed
superfields are still chiral. Furthermore, one may show that 
the fermionic derivatives are invariant under $SU(1,1)$ transformations,
\be
D_\a' = D_\a,
\ee
when acting on functions of the dilaton superfield.
This can be shown by starting from (\ref{ch}) and acting with an 
$SU(1,1)$ (or $SL(2,R)$) transformation. Since we know how all
superfields transform,  it is straightfoward to determine how
the fermionic derivatives transform.

\subsection{Projections}

All fermionic derivatives of the chiral superfield  
$V$ as well as their
evaluation at $\th=0$ can be computed using the   
information in the previous section.  The first derivative of   
$V$, $D_\a V$, is given in (\ref{ch}). To compute $D_\a D_\b V$,   
we need $D_\a \L_\b$ and $D_\a U$. These are given in (\ref{DL})   
and (\ref{ch}). At the next level, $D_\a D_\b D_\g V$, we need   
in addition the derivative of $F_{abc}$. This is given in   
(\ref{dF}). In order to compute the fourth derivative of the   
superfield $V$, we now need the derivatives of $\L_\a$,   
$F^*_{abc}$, $P_a$ and $\Psi_{ab}{}^\a$, all of which   
are given in the previous section (see equations   
(\ref{DL})-(\ref{bDL})-(\ref{dbF})-(\ref{DFDPDPsi})).   
Notice that up to this order, new component fields were   
involved at every order: the dilatino appears in   
$D V$, $F_{abc}$ appears at $D^2 V$, the gravitino field
strength appears at $D^3 V$ and the Riemann tensor and $F_5$
appear at order $D^4 V$.

In order to proceed from here, we need the derivatives of all quantities
appearing above.  The only new derivatives we have to evaluate are   
the derivative of the curvature $R_{abcd}$, $\bar P_a$ and of the   
torsions in \eq{DFDPDPsi}. The
components of the torsion involved in the above expression are
$T_{a\a}{}^\b$, $T_{a\a}{}^{\bar\b}$ and $T_{a\bar\a}{}^\b$.
These are given entirely
in terms of the superfields $\L$, $\bar\L$, $Z_{abcde}$, $F_{abc}$ and   
$F^*_{abc}$, where $Z_{abcde}$ is defined in (\ref{Z}).
So to differentiate \eq{DFDPDPsi}, we only need to
determine $D_\a \L^*_\a$, $D_\a Z_{abcde}$, $D_\a R_{abcd}$ and   
$D_\a\bar P_a$. All of these are given in the previous   
section. At the next level the only new derivative that we encounter is
$D_\a\Psi_{ab}{}^{\bar\a}$ and this is given in  
(\ref{dPsib}).

In the previous section we gave the $\th=0$ components and   
first covariant derivative of all fields.  We can therefore
determine all fermionic derivatives of the superfield $V$.
Furthermore, the evaluation of the derivatives of $V$ at $\th=0$ 
is straightforward but tedious given the results in the previous section. 
One should  remember that the evaluation of the field strengths at $\th=0$
gives rise to supercovariant objects.

To illustrate the procedure we discuss here the computation   
of the first four projections of $V$. This will be of use   
in section \ref{sl2} where we compute the components   
of the supepotential. Following the procedure discussed   
above we get
\bea
V| &=& v \\
D_\a V| &=& - 2 u \l_\a \\
D_{[\a} D_{\b]} V| &=& {i \over 12} u \g_{\a \b}^{abc} \hat{f}_{abc}   
\label{d2v} \\
D_{[\g} D_\b D_{\a]} V| &=& \frac{i}{12}  u \g^{abc}_{\b \a}
\left\{
-\frac{1}{32} \left( \g_{abcdef} \hat{f}^{* def}
+ 3 \hat{f}^*_{ [a}{}^{de}\g_{bc]de}
+ 52 \hat{f}^*_{ [ab}{}^{d} \g_{c]d}
+ 28 \hat{f}^*_{ abc} \right)_\g{}^{\e} \l_\e \right.
\nonu
&+& \left. 3 \hat{p}_{[a} (\g_{bc]})_\g{}^\e \l^*_e
+ 3 i (\g_a)_{\g \e} \hat{\psi}_{bc}^\e \right\}.
\eea
These formulas are exact in that they contain all bosonic
and fermionic terms. As we discussed, this procedure can be continued   
till all projections are obtained. The number of terms involved in the   
computation, however, grows as we go up in level. Since the computation   
is algorithmic, it can presumably be computerized. In the remaining   
of this section we compute the bosonic part of the fourth
projection. The computation proceeds by taking the fermionic   
derivative of $D^3 V$. Keeping only terms that   
contribute purely bosonic terms we get
\bea \label{d4v}
D_{[\d} D_{\g} D_{\b} D_{\a]} V &=& \frac{i}{12} U \g^{abc}_{\b \a}
\left\{ -\frac{1}{32} \left( \g_{abcdef} F^{* def}
+ 3 F^*_{ [a}{}^{de}\g_{bc]de}
+ 52 F^*_{ [ab}{}^{d} \g_{c]d}
+ 28 F^*_{ abc} \right)_\g{}^{\e} D_\d \L_\e \right.
\nonu
&+& \left. 3 P_{[a} (\g_{bc]})_\g{}^\e D_\d \L^*_\e
+ 3 i (\g_a)_{\g \e} D_\d \Psi_{bc}^\e \right\}.    
\eea
Using the expression for $D_\d \L_\e$, $D_\d \L^*_\e$ and
$D_\d \Psi_{bc}^\e$ given in the previous section   
we obtain a number of bosonic terms.
In particular, the $D_\d \Psi_{bc}^\e$ term gives terms     
linear $R$ and $D F_5$. These are the terms that are present   
in the linearized superfield of Howe and West.
This part is given by
\bea
D_{[\d} D_{\g} D_{\b} D_{\a]}  V|_{\sm{linear}} &=&    
\frac{1}{16} u (\g^{abc})_{\b \a} (\g_a{}^{de})_{\g \d}
r_{bcde} - \frac{i}{96} u (\g^{abc})_{\b \a} (\g^{def})_{\g \d} D_b
f_{acdef} \nn
&=&   \frac{1}{16} u   
(\g^{abc})_{\b \a} (\g^{def})_{\g \d}
(g_{ad} c_{bcef}     
- \frac{i}{6} D_b f_{acdef})~.     
\eea
Here $c_{bcde}$ is the Weyl tensor, and in passing to the   
second equality we used the Fierz identity (\ref{Fierzing})
to show that only the Weyl tensor contributes.   
  
There are further contributions, however, that are not   
captured by the linearized superfields. They are proportional
to $f^*_3 f_3$ and $f_5 f_5$. The latter give
\be \label{f5nl}
D_{[\d} D_{\g} D_{\b} D_{\a]} V|_{f_5 f_5} =    
-\frac{u}{1536}  
\g^{abc}_{[\b \a} \g^{def}_{\g \d]}
(3 f_{bafmn} f_{ced}{}^{mn}  
- f_{abcmn} f_{def}{}^{mn})~.
\ee
The computation leading to this term is elaborate and is given   
in appendix \ref{NL}. The $f^*_3 f_3$ terms can also be computed   
straightforwardly, but we shall not present them here.

To summarize, we obtained
\be
D_{[\d} D_{\g} D_{\b} D_{\a]} V| =   
u \g^{abc}_{[\b\a}\g^{def}_{\g \d]} \car_{abcdef}
\ee
where   
\be \label{car}
\car_{abcdef} =\frac{1}{16}(g_{ad} c_{bcef}     
- \frac{i}{6} D_b f_{acdef})
-\frac{1}{1536} (3 f_{bafmn} f_{ced}{}^{mn} 
- f_{abcmn} f_{def}{}^{mn})
 + f^*_3 f_3 {\rm\ terms}~.
\ee

\section{Obstruction to a supersymmetric action} \label{saction}
\setcounter{equation}{0}

In supergravity theories the measure $e=\det e_m^a$ transforms under   
supersymmetry. To construct actions one needs an appropriate
supersymmetric measure. For unconstrained superfields the
supersymmetric measure is given by the superdeterminant of
the supervielbein, sdet$\,E$. Chiral superfields, however, are only integrated
over half of superspace, and sdet$\,E$ is not the correct density.
In four dimensions the chiral measure is a chiral superfield
whose lowest component is $e$. In our case, however, the superfield $\D$ whose
lowest component is $e$ cannot be a chiral scalar superfield.
As discussed in the previous section, chiral superfields   
should satisfy (\ref{chL}). From the supersymmetry rules
one obtains $D_\a \Delta| \sim (\g^b \psi_b)_\a$, so
(\ref{chL}) is not satisfied.  We will proceed by systematically
analyzing the constraints imposed on the measure by supersymmetry. 

We consider the following action
\be \label{supera}
S = \int d^{10}x \,d^{16} \Theta\, \D\, W[V,U^*] + \mbox{c.c.}
\ee
where the superpotential   
$W[V,U^*]$ is an arbitrary function of the chiral superfields $V, U^*$
but not of their derivatives or complex conjugate superfields, i.e.
$W[V,U^*]$ is itself a chiral superfield.
By definition,   
\be
d^{16} \Theta= D^{16} \equiv
\frac{1}{16!} \e^{\a_1..\a_{16}} D_{\a_1} \ldots D_{\a_{16}}~.  \label{defD}
\ee
The reason for considering (\ref{supera}) is that an action
of this form was argued to capture the interactions of the 
effective action at the linearized level, see \cite{Green:1999qt} and 
references therein. Furthermore, a scalar superpotential term is one of 
the simplest interaction terms one may consider. There are other 
possibilities one may consider such as considering a non-scalar 
superpotential.

The action should be gauge invariant. Each $D_\a$ has charge
$-1/2$ under the local $U(1)$ symmetry. This implies that $W$   
must have charge +8 in order for the action to be $U(1)$
invariant. This can be achieved by setting
\be
W= (U^*)^8 \tilde{W}(A)
\label{superpotential}\ee
where $A$ is the gauge invariant dilaton superfield
(since both $A$ and $T$ are gauge invariant one may    
consider $\tilde{W}$ as either a function
of $A$ or $T$). The factor $U^*$ may be considered
as a $U(1)$ compensator. Gauge fixing the $U(1)$ symmetry
amounts to setting $U^*$ equal to some function of   
$A$.

We shall determine the constraints imposed by supersymmetry 
on $\D$ and analyze whether there is a $\D$ such that the 
action is supersymmetric. 
By definition, the chiral measure $\D$ is a superfield whose
lowest component is the determinant of the vielbein
\be \label{Dlea}
\Delta| = {\rm det}  \,e_m{}^a = e~.
\ee
The idea now is to determine the remaining projections 
by systematically arranging that the supersymmetry variation 
of the action vanishes. We shall see that IIB superspace
allows for a precise formulation of the problem along these
lines.
 
Integrating out the $\Theta$ one obtains the following   
component action
\bea \label{action}
S &=& \int d^{10} x\,  \e^{\a_1..\a_{16}}   
\sum_{n=0}^{16} {1 \over n! (16 -n)!}    
D_{\a_1} ..D_{\a_{n}} \D|\, D_{\a_{n+1}}...D_{\a_{16}} W| \nn
&=& \int d^{10} x \sum_{n=0}^{16} {1 \over n!}   
D_{\a_1} ..D_{\a_{n}} \D|\, D^{16-n,\a_1 ..\a_n} W|
\eea
where we have introduced the notation
\be \label{def}
D^{16-n,\a_1 ..\a_n} W = {1 \over (16-n)!}\, \e^{\a_1..\a_{16}}   
D_{\a_{n+1}}...D_{\a_{16}} W~.
\ee
One would like to fix the higher projections   
of $\Delta$ such that this action is supersymmetric for any   
superpotential $W$. Notice that because of the $\e$-symbol only   
fully antisymmetrized projections of $\D$ enter in the action.

Invariance of the action under supersymmetry requires
\be \label{daction}
\d S = \int d^{10} x   
\sum_{n=0}^{16}   
\frac{1}{n!}  \left(   
\d D_{\a_1} ..D_{\a_{n}} \D| D^{16-n,\a_{1}..\a_{n}} W|   
+ D_{\a_1} ..D_{\a_{n}} \D| \d D^{16-n,\a_{1}..\a_{n}} W| \right)=0.
\ee
Since we consider consider arbitrary superpotential $W$ 
the terms $D^n W|$ are linearly independent and one can investigate 
their cancellation separetely. Furthermore, this can be 
done systematically by starting from the
term with the highest number of derivatives,   
$D^{16} W|$,  and then moving to $D^{15} W|$ terms,
etc. For each $D^n W|, n=1,..,16$, supersymmetry implies
two conditions: one for the terms that are proportional
to $\z$ and another for the ones proportional to $\z^*$. 
We show below that the conditions proportional 
to $\z$ uniquely determine all projections of the chiral
measure $\D$. This leaves 16 more conditions to be checked: the 
ones proportional to $\z^* D^n W|$. These conditions should be 
satisfied  automatically for (\ref{supera}) to be supersymmetric. Furthermore, 
since the coefficients of  $\z^* D^n W|$  are field dependent,
these conditions further split into a number of independent conditions: 
one for each independent structure.

The supersymmetry variation of $D^{16-n} W|$ is given by
\be \label{susyW}
\d D^{16-n,\a_1..\a_{n}} W| =
{1 \over (16-n)!} \e^{\a_1 .. \a_{16}}
(\z^\a D_\a D_{\a_{n+1}} ..D_{\a_{16}} W|
- \z^{*\a} D^*_{\a} D_{\a_{n+1}} ..D_{\a_{16}} W| )
\ee
Inspection of the supersymmetry algebra reveals that   
(anti) commutators of covariant derivatives cannot   
increase the number of $D_\a$ derivatives acting on $W$.
It follows that the schematic form of the variation is   
\be \label{sW}
\d D^{16-n} W| \sim \z \left[ D^{17-n} W|   
+ O(D^{15-n} W|) \right] + \z^* O(\psi D^{16-n} W|)~.
\ee
Let us show this in some detail.
Consider first the terms proportional to $\z^{*\a}$.
Anticommuting $D^*_\a$ to the right where it annihilates
$W$ yields $D_a D^{15-n}W|$ plus other terms of order $D^{15-n}W|$.
The term $D_a D^{15-n}W|$ is actually of order $D^{16-n} W|$.
To see this, use (\ref{flcu}) to convert 
the flat index in $D_a$ to a curved one. This yields
a term $\psi D^{16-n} W|$ plus terms of lower order.
Thus the $\z^\a{}^*$ terms are of order $D^{16-n} W|$.
Let us now discuss the $\z^\a$ terms.
Antisymmetrizing all derivatives we get $ D^{17-n} W|$.
The antisymmetrization involves anticommutators
$\{D_{\a_i}, D_{\a_j}\}$, and the corresponding
curvature terms are of order $D^{15-n} W|$.
The torsion term yields $D^*_\a D^{15-n} W|$, which can be 
analyzed as the term  $D^*_\a D^{16-n} W|$, and is also
of order $D^{15-n} W|$. We conclude that the supersymmetry variation
of $D^{16-n}W|$ is of the form (\ref{sW}).
 
It follows from (\ref{sW}) that one can iteratively 
determine all projections of $\D$ by arranging that the 
$\z$ terms in the variation of the action cancel. In particular,
the cancellation of the terms of order $D^{17-n} W|$ determine
the projection $D_n \D|$. We now prove this inductively.
The $n=1$ case will be shown to hold in the next subsection.
Let us assume that the statement holds true for all $n<k$,
i.e. we assume that we determined all $D_n \D|, n<k$, by arranging 
for the cancellation of the terms proportional to $D^{17-n} W|, n<k$. 
Let us now consider the $n=k$ case. From (\ref{daction}) we get 
\be \label{dnd}
\d_\z S = \int d^{10} x 
\left(\d_\z D^{k-1} \D|  + 
\z D^k \D|  + \z f(D_n \D|) \right) 
D^{17-k} W| + O(D^{16-k} W|)
\ee
where we supress indices and numerical factors, 
$\d_\z$ denotes a supersymmetry variation with only  
$\z$ terms taken into account, and $f(D_n \D|)$ denotes the terms 
that originate from manipulations of higher order terms. 
To be more explicit, recall that 
the supersymmetry variation $\delta D^{16-n} W$ in (\ref{sW})
involves terms proportional to $D^{16-l} W|, l > n$.
Thus the terms $D_n \D (\delta D^{16-n} W|), n < k$ in the 
variation of the action can contribute term proportional
to $\z D^{17-k} W|$. These terms depend on $D_n \D, n<k$, 
which are known by the induction hypothesis and are denoted
by $f(D_n \D|)$. It follows that by setting 
\be
\z D^k \D| \sim (\d_\z D^{k-1} \D)| + \z f(D_n \D|)
\ee
the $\z D^{17-k} W|$ terms vanish, as advertised. 
This finishes the inductive proof that the cancellation of 
$\z$ terms uniquely determine all components of $\D$.
In the next subsection we will determine the exact form,
including coefficients, of $D_\a \D|$ and 
$D_{[\a} D_{\b]} \D|$ (up to certain fermion bilinears for
the latter).

We are now left with the $\z^*$ terms in the supersymmetry 
variation. For the action to be supersymmetric, these
terms should vanish automatically. We shall see that 
these terms cancel at leading order and at a linear level
at the subleading order, but there is an obstruction 
which is non-linear in the fields in subleading order\footnote{
In the original version of this paper, we only considered 
the linear subleading terms. We are grateful to Nathan Berkovits and Paul Howe
for informing us that the $\l \l^* D^{15} W|$ terms present an obstruction.}.

\subsection{Obstruction at subleading order}

We now discuss in detail the determination of the first 
two projections of $\D$ and the obstruction to the 
existence of the chiral measure.
Using (\ref{Dlea}) and keeping terms up to order $D^{15} W|$ 
in the supersymmetry variation of the action, we find
\bea
\d S &=& \int d^{10} x \left[ \d e D^{16} W| + e (\d D^{16} W|)
+  
e \left(  
(- \d D^{15,\a} W| D_{\a} \Delta| \right. \right.  \nn
&&  \hspace{-1cm}
\left. \left. - D^{15,\a} W| \d D_{\a} \Delta|)
+ \frac{1}{4} \d D^{14,\a_1 \a_{2}} W|
[D_{\a_{1}}, D_{\a_{2}}] \Delta|
+ \ldots \right) \right] =0 \label{varact}
\eea
The variation of $e$ can be computed from the supersymmetry variation
of the vierbein 
\be
\d e = -ie ( (\zeta \gamma \psi^*) + (\zeta^* \g \psi) )~.
\ee
The variation of the $D^{16} W|$ term is given by
\be
\d D^{16} W|=\frac{1}{16!} \e^{\a_1 ...\a_{16}}
(\zeta^{\a} D_{\a} - \zeta^{* \a} D^*_{\a} ) D_{\a_1} ... D_{\a_{16}} W|~.
\ee
Let us consider separately the $\z$ terms and the $\z^*$ terms.
To manipulate the $\z^*$ terms we need to compute the commutator
$[D^{*}_{\a}, D^{16}] W|$. Using the supersymmetry algebra   
we obtain
\bea 
[D^{*}_{\a}, D^{16}] W|&=&  \frac{1}{16!} \e^{\a_1 ...\a_{16}}   
\left[ -16 T_{\bar{\a} \a_1}^c D_c D_{\a_2} ..D_{\a_{16}} W|   
-\frac{15 \cdot 16}{2} \left( T_{\bar{\a} \a_2}^c   
T_{\a_1 c}^{\d} D_\d D_{\a_3} .. D_{\a_{16}} W| \right. \right.\nn
&+&  \hspace{-.3cm} \left. \left. R_{\bar{\a} \a_1 \a_2}{}^{\d}   
D_{\d} D_{\a_3} ..D_{\a_{16}} W| \right) 
+ i {17 \cdot 16 \over 2} M_{\a_1 \bar{\a}} D_{\a_2} .. D_{\a_{16}} W| 
\right] + O[D^{14} W|], 
\eea
where the curvature terms originate in the anticommutators,
$\{D^*_{\a}, D_{\a_p} \}$, and the double torsion terms arise
in the process of commuting $D_c$ to the left.
We now convert the $D_c$ derivative to a curved space
derivative, $D_m$, using $D_c=E_c^M D_M$,
\bea \label{d*1} 
[D^{*}_{\a}, D^{16}] W|&=&
T_{\bar{\a} \d}^c \psi_c^{\d} D^{16} W| -\left(e_c^m T_{\bar{\a} \b}^c D_m   
\right. \\
&&\hspace{-1cm}\left. - i {17 \over 2} M_{\b \bar{\a}}
+\frac{1}{2} ( T_{\bar{\a} \g}^c T_{\b c}^{\g} 
- T_{\bar{\a} \b}^c T_{\g c}^\g   
- R_{\bar{\a} \g \b}{}^{\g}) + O[(\psi^* \psi)] \right)
D^{\b, 15} W| 
+ O[D^{14} W|]. \nonumber 
\eea

In the $\z$ terms we proceed by fully antisymmetrizing the   
$D_\a$ derivative with the rest of the derivatives. Since the   
index $\a$ takes only 16 values, the fully antisymmetric product
is identically equal to zero. We therefore obtain
\be
\e^{\a_1 ...\a_{16}} D_\a D_{\a_1}..D_{\a_{16}} W =
\e^{\a_1 ...\a_{16}} \sum_{p=0}^{15}   
(-1)^{p} (16-p) D_{\a_1} .. D_{\a_p} \{D_{\a_{p+1}}, D_\a \} D_{\a_{p+2}}
..D_{\a_{16}} W~.
\ee
We now use the superalgebra
\be
\{D_{\a_{p}}, D_\a \} = - T_{\a_p \a}^{\bar{\b}} D^*_\b   
+ \half R_{\a_p \a \b}{}^\g L_{\g}{}^\b~.   
\ee   
The terms with $D^*_\b$ can be manipulated in a way similar to
(\ref{d*1}). They give rise to terms of order $\psi \l^* D^{15} W|$.
The rest yields,
\be
D_\a D^{16} W| = {1 \over 3} R_{\a \b \g}{}^\b D^{15,\g} W|
+ O(\psi \z^* D^{15} W|).
\ee
Similar manipulations yield
\be
\frac{1}{15!} \e^{\a_1 .. \a_{16}} \d D_{\a_1} .. D_{\a_{15}} W| D_{\a_{16}}
\Delta|
= - (\zeta D \Delta|) D^{16} W| + O(\psi \l^* D^{15} W|)~.   
\ee

At this point we have all contributions to order $D^{16} W|$.
Keeping terms of leading order only we obtain
\bea   
\d S &=& \int d^{10}x \left[\z^\a   
(-i e \g^c_{\a \b} \psi^{* \b}_c- D_\a \D|)  D^{16} W| \right.  \nn
&& \left.
- \z^{* \a}([D_\a^*, D^{16}]W| + i e \g^c_{\a \b} \psi_c^\b D^{16} W| )   
+ O(D^{15} W|) \right]. \label{w16}
\eea
The cancellation of the $\z$-terms uniquely fixes the first projection of
$\D$,
\be
D_\a \D| = - i e \g^c_{\a \b} \psi_c^{* \b}~. \label{fpdelta}
\ee
The $\z^*$ terms at leading order vanish by themselves upon
using the leading order term in (\ref{d*1}).

We next move to terms of order $D^{15} W|$. We will keep
terms that are linear in the fields and from the fermion bilinears 
only the terms proportional to $\l \l^*$. We now need
to compute   
\bea
\d D_{\b} \Delta| D^{\b, 15} W| &=& T_{\b \bar{\g}}^c \d (e \psi^{* \g}_c)
D^{\b, 15} W| \nn
& =& T_{\b \bar{\g}}^c e   
\left(e_c^m D_m \zeta^{* \g}  + \zeta^{\a} T_{\a c}^{\bar{\g}} -   
\zeta^{* \a} T^{\bar{\g}}_{\bar{\a} c} \right) D^{\b, 15} W| \nn
&&+ O(\psi^* \psi D^{15} W| ) + O(D^{14} W|)~,
\eea
where we found it useful to use (\ref{susy2}) for
$\delta \psi^{* \g}_c$ rather than substituting the expression from   
appendix \ref{st}. Furthermore,
\be
\frac{1}{4 \cdot 14!} \e^{\a_1 .. \a_{16}}   
\d D_{\a_1 ...\a_{14}} W| [D_{\a_{15}}, D_{\a_{16}}] \Delta| =
- \frac{1}{2} \zeta^{\a} [D_{\a}, D_{\b}] \Delta| D^{\b, 15} W| +
O(D^{14} W|)
\ee

Summing up all contributions we obtain
\bea  \label{ll}
\d S_{15} &=& \int d^{10} x e \left\{
D_m \left(e_c^m T^c_{\bar{\b} \a} \zeta^{* \a} D^{\b, 15} W|\right)
\right.  \nn
&& - \zeta^{\a} \left[ i 
\g^{c}_{\b \g} T_{\a c}^{\bar{\g}} - \frac{1}{3} R_{\a \g \b}{}^{\g}
+ \frac{1}{2} [D_{\a}, D_{\b}] \Delta| \right] D^{\b, 15} W| \\
&&\left. + \zeta^{*\a} \left[-i {17 \over 2} M_{\b \bar{\a}}+
\frac{1}{2} ( T_{\bar{\a} \g}^c T_{\b c}^{\g} 
- T_{\bar{\a} \b}^c T_{\g c}^\g - R_{\bar{\a} \g \b}{}^{\g}  )
- T^c_{\b \bar{\g}} T_{\bar{\a} c}^{\bar{\g}}   
\right] D^{\b, 15} W| \right\}  \nonumber
\eea
The total derivative term originates from (\ref{d*1}) and the
$D_m \z^*$ term in $\delta \psi^{* \g}_c$.

Requiring that the $\z$-terms cancel determines the
projection $[D_\a, D_\b] \D|$,
\be
i e
\g^{c}_{\b \g} T_{\a c}^{\bar{\g}} - e \frac{1}{3} R_{\a \g \b}{}^{\g}
+ \frac{1}{2} [D_{\a}, D_{\b}] \Delta| =0~.
\ee
Using the explicit formulas for the curvature and torsions we obtain
\be
i e \g^{c}_{\b \g} T_{\a c}^{\bar{\g}} =
\frac{1}{24} i e \g^{abc}{}_{\a \b} f^*_{abc}  \qquad
R_{\a \g \b}{}^{\g} =
\frac{1}{4} i e \g^{abc}{}_{\a \b} f^*_{abc}
\ee
which leads to  
\be
[D_{\a}, D_{\b}] \Delta| = \frac{1}{12} i e \g^{abc}{}_{\a \b} f^*_{abc}
+ O[\psi \psi^*,\l^* \psi] ~.
\ee

We process the $\z^*$ terms by using the Bianchi identities.
They imply
\bea
&&I^{(1)}_{\bar{\a} \g \b}{}^{\g} = 0 \quad \Rightarrow \quad 
R_{\bar{\a} \g \b}{}^{\g} = - 17 i M_{\b \bar{\a}}
- T_{\bar{\a} \g}^c T_{\b c}^\g - T_{\b \bar{\a}}^c T_{\g c}^\g
- T_{\g \b}^{\bar{\d}} T_{\bar{\d} \bar{\a}}^\g \\
&&I^{(1)}_{\a \bar{\b} c}{}^{c}=0 \quad \Rightarrow \quad
T_{\bar{\a} c}^{\bar{\g}} T_{\bar{\g} \b}^c -
T_{\b c}^{\g} T_{\g \bar{\a}}^c =0
\eea
where we used the notation introduced in (\ref{b1}).
Inserting these expression in (\ref{ll}) we find that all terms
but one cancel out and we end up with
\be \label{rem}
\d S_{15} = \half \int d^{10} x 
e \zeta^{*\a} T_{\bar{\a} \bar{\d}}^\g T_{\g \b}^{\bar{\d}}  D^{\b, 15} W|
\ee
Recall that we showed in section 4 that 
$T_{\bar{\a} \bar{\d}}^\g \neq 0$ is the obstruction
for the existence of chiral superfields. Here we find that it is also
the obstruction for the existence of a chiral measure.

In type IIB superspace $T_{\bar{\a} \bar{\d}}^\g$ is non-zero.
Nevertheless, the dilaton superfield $V$ exists because the 
torsion coefficient is such that $T_{\bar{\a} \bar{\d}}^\g D_\g V=0$. 
The superpotential is a function of $V$, so one should check 
that the obstruction does not vanish because of special properties
of the dilaton superfield. Using the notation we introduce in 
(\ref{fv}) one finds,
\be 
D^{\b, 15} W| = 
{1 \over 15!} \e^{\b \a_1...\a_{15}} D_{\a_1} V...D_{\a_{15}} V F^{(15)}(V)|
+ ...
\ee
where the dots indicate terms with $F^{(n)}(v), n<15$. Since (\ref{rem})
should be valid for any superpotential, the terms proportional to 
different $F^{(n)}(v)$ should vanish separately. Let us consider the 
term proportional to $F^{(15)}(v)$. Notice also that such a
term is present only in $D^{15} W|$, so there cannot be
any cancellations involving terms with $D^{n} W|$, $n<15$.
Using (3.30) we get
\be \label{rem1}
\d S_{15, F^{(15)}} = \half \int d^{10} x 
e \zeta^{*\a} (2 u)^{15} 
T_{\bar{\a} \bar{\d}}^\g T_{\g \b}^{\bar{\d}} \l^{15,\b} F^{(15)}(v) 
\ee
Using the explicit form of the torsion coefficients we finally get
\be 
\d S_{15, F^{(15)}} = - 77 \int d^{10} x 
e (\zeta^{*} \l^*) \l^{16} (2 u)^{15} F^{(15)}(v)
\ee
which is non-zero. We conclude that the chiral measure does not 
exist and the action (\ref{supera}) is not supersymmetric. This computation
still leaves the possibility that the measure exists at the 
linearized level. To check that, one needs to analyze the 
terms $D^nW|$, with $n \leq 14$. 

In this section we investigated the existence of a scalar measure:
all supercovariant derivatives in (\ref{defD})
are completely antisymmetrized. Terms similar
to the ones in (\ref{rem1}) will be generated if one relaxes the 
full antisymmetrization. In this case, however, the measure $d^{16} \Theta$
would have free indices, and therefore the superpotential should also
carry indices. Such terms are also generated if the chirality constraint
on the superpotential is relaxed. Perhaps there is a simple modification 
of the construction presented in this paper that will be supersymmetric
and hopefully be related to the $R^4$ term that appears 
in string theory.

In the next section we  consider the component 
form of the action in (\ref{supera}) with $\D$ determined by 
the cancellation of $\z$-terms, as described in this section.
This action is not supersymmetric but as we shall see
its properties are rather intriguing.

\section{Components and $SL(2,Z)$ symmetry} \label{sl2}
\setcounter{equation}{0}

\subsection{The superpotential in components}

We discuss in this section the computation of the superpotential
in components. The computation consists of evaluating at   
$\th=0$ the terms in (\ref{action}). Since we know
all components of $V$ and we showed how to determine the
components of $\D$ (and explicitly determined the first two), 
it is straightforward but tedious
to obtain all components. We will discuss in detail the computation   
of the terms proportional to $\l^{16}$, $\psi^* \l^{15}$   
and $r^4$.   

First, notice that we can solve (\ref{det}) to express $U^*$   
in terms of $V$,
\be
U^* = A + B V~,
\ee
where $A=1/U$ and $B=V^*/U$. Since $A$ and $B$ are annihilated by
$D$, they can be considered as constants for the purpose of evaluating
the fermionic integral. We now define
\be \label{fv}
F(V) = W[V,U^*{=}A+ B V] = (A+ B V)^8 \tilde{W}\left[{V \over A+ B V}\right],
\ee
where we used \eq{superpotential}. With this definition we now schematically 
have
\be   
S \sim \int d^{10} x \sum_{n=0}^{16} \sum_{k=1}^{16-n} F^{(k)} (v)
D^n \D| \sum_{\sum n_i=16-n}   
D^{n_1} V| \cdots D^{n_k} V|
\ee
where $F^{(n)} = \pa^n F/ \pa V^n$,   
and we suppress combinatorial factors (which however will be taken   
into account below).

We want to compute specific terms in the component expansion
of the superpotential, namely the $\l^{16}$, $\psi^* \l^{15}$   
and $r^4$.   
To this end we need to find a way to obtain which of the   
projections contribute. To do this we will use the local   
$U(1)$ symmetry and dimensional analysis. Let us first discuss the $r^4$ term.
The Riemann tensor $r$ is neutral under the local $U(1)$.
Therefore each of the projections contributing to $r^4$   
should also be neutral under the local $U(1)$.  From the analysis in  
section (\ref{dilsec}) the form  of the projections is
\be
D^n V| \sim u g~,   
\ee
where $g$ is a function of the fields that does not depend on $u$ or $v^*$.
Now, from table 1 we see that
the covariant derivative has $U(1)$ charge $-1/2$, $v$ charge $1$
and $u$ charge $-1$. This implies that the function $g$   
has charge $2-n/2$. It follows that only the $n=4$ projection
is $U(1)$ neutral. A similar argument applies to the projections
of $\D$. In this case only the leading component is $U(1)$
neutral. Notice that terms of the from $D^4 \D$ are excluded
by a combination of dimensional and U(1) analysis.
We therefore obtain that the $r^4$ term is given by
\bea \label{r4com}
S_{r^4} &=& \int d^{10} x\, e {1 \over (4!)^5}\, F^{(4)}(v) (D^4 V)^4 \nn
&=&\int d^{10} x\, e\, {1 \over (4!)^5}\, u^4 F^{(4)}(v) \car^4
\eea
where $\car$ is given in (\ref{car}), and the exact  
index contractions are given in appendix \ref{r4}, see (\ref{SR4}).
In the same appendix we show that index contraction of the $c^4$
term in $\car^4$ is the same   
as that of the $c^4$ term of string theory \cite{GW}.

We thus see that the superpotential  contain terms of the form   
$c^k (Df_5)^l (f_5)^m$, for appropriate values of $k,l,m$.
One may extend the computation described in appendix \ref{r4}
to obtain the exact index contractions, but we shall not do this   
here.

Let us now analyze the function $F^{(4)}(v)$. Using the   
definition in (\ref{fv}) we obtain
\be
u^4 F^{(4)}(v) = \sum_{n=0}^4 c_n^{(0)} (u^* v^*)^n \tilde{W}^{(4-n)}(a)
\ee
where $\tilde{W}^{(n)}=\pa^n \tilde{W}/\pa a^n$ and the combinatorial 
coefficients are given by
\be \label{coef}
c_0^{(0)}=1, \quad c_1^{(0)}=20, \quad c_2^{(0)}=180, \quad c_3^{(0)}=840,
\quad c_4^{(0)}=1680.
\ee

Let us now consider the $SL(2,R)$ Laplacian,
\be
\nabla^2 = 4 \tau_2^2 {\pa \over \pa \tau}  {\pa \over \pa \tau^*}   
= (1-a a^*)^2 {\pa \over \pa a}  {\pa \over \pa a^*}    
\ee
where we also express it in terms of $SU(1,1)$ physical scalars.   
An easy computation yields,
\be \label{naw}
\nabla^2 ((u^* v^*)^n \tilde{W}^{(m)}) =
n(n+1) (u^* v^*)^n \tilde{W}^{(m)} + n (u^* v^*)^{n-1} \tilde{W}^{(m+1)}.
\ee
Using this result we then obtain,
\be \label{20}
\nabla^2 (u^4 F^{(4)}(v)) = 20 (u^4 F^{(4)}(v))~,
\ee
where the exact values of the coefficients were crucial for  
$u^4 F^{(4)}(v)$
to be an eigenfunction of the Laplacian. Looking through the   
computation we see that the eigenvalue 20 is basically due to the   
fact that the $r^4$ term comes from the fourth power of the   
fourth projection of $V$.   

The $c^4$ coupling receives a contribution from the   
complex conjugate of the superpotential term as well. The   
analysis is exactly the same (it is the cc analysis   
of what we presented). We thus finally define
\be
t^{(0,0)}(\t, \t^*) = {6^4 \over (4!)^5 16^4}
(u^4 F^{(4)}(v) + u^*{}^4 F^*{}^{(4)}(v^*)).
\label{t00}\ee
The factors $1/16^4$ originate from a similar factor 
in (\ref{car}) and the factor $6^4$ from the manipulations
described in appendix E.
We will shortly show that imposing $SL(2,Z)$ symmetry implies
that $t^{(0,0)}(\t, \t^*)$ is equal to the non-holomorphic
Eisenstein series $E_5$, but before we do this we will examine
the $\l^{16}$ and $\l^{15} \psi^*$ terms and the $SL(2,R)$   
properties of their coefficients.

Similar considerations as the ones above show that the
only term that can contribute to the $\l^{16}$ term   
is $(DV|)^{16}$, and that the measure cannot contribute.
We thus have
\bea
S_{\l^{16}} &=& \int d^{10}x\, e\, F^{(16)} {1 \over 16!}   
\e^{\a_1...\a_{16}} D_{\a_1} V| \cdots D_{\a_{16}} V| \nn
&=& \int d^{10}x\, e\, 2^{16} u^{16} F^{(16)} \l^{16}
\eea
where we define\footnote{Here we follow the conventions
in \cite{GrSe}. This definition differs by a sign when $n$ is odd from 
the similar definition in (\ref{def}).} 
\be
(\l^{n})_{\a_{n+1}..\a_{16}} = {1 \over n!} \e_{\a_1..\a_{16}}   
\l^{\a_1} \cdots \l^{a_n}~.
\ee
Let us call the coefficient of $\l^{16}$, $t^{(12,-12)}$ (the reason
for the terminology will become apparent later). Then   
\be \label{t12}
t^{(12,-12)} = 2^{16} u^{16} F^{(16)}~.
\ee

Let us now consider the terms $\psi^* \l^{15}$. There are two sources
of such terms. One comes from $\D| D^2 V| (DV|)^{14}$
and another receives a contribution from the measure
$(D \D)| (D V|)^{15}$. The former contributes because
$D^2 V|$ is proportional to the supercovariant   
field strength $\hat{f}_3$ (\ref{d2v}), and the latter contains
a $\psi^* \l$ term, see (\ref{scov}). This contribution is also   
discussed by \cite{GrSe}, but we get an additional term
from the measure. Combing the two we obtain
\be   
S_{\psi^* \l^{15}} = \int d^{10} x\,  \left(-i e 2^{18} u^{15} F^{(15)}\right)
(\psi^*_c \g^c \l^{15})~.
\ee
Let us define
\be \label{t11}
t^{(11,-11)} = -i 2^{18} u^{15} F^{(15)}
\ee
to be the coefficient of the $\psi^* \l^{15}$ term. Comparing   
(\ref{t12}) and (\ref{t11}) we see that they satisfy
\be \label{constr}
u {\pa \over \pa v} t^{(11,-11)} = - 4 i t^{(12,-12)}~.
\ee

The coefficients $t^{(12,-12)}$ and $t^{(11,-11)}$ are analogous to the   
coefficients $f^{(12,-12)}$ and $f^{(11,-11)}$ introduced by Green   
and Sethi in \cite{GrSe} (but we view   
$f^{(11,-11)}$ as the coefficient of $\psi^* \l^{15}$ rather than   
$\hat{f}_3 \l^{14}$, i.e. we view (3.3) of \cite{GrSe} rather than (3.1) as   
the starting point of their analysis. As discussed above, the coefficient
of $\psi^* \l^{15}$ receives a contribution from the measure as well).
A supersymmetry analysis that uses only the lowest order supersymmetry   
rules leads them to the constraint (3.5) which after rescaling their
$f^{(11,-11)}$ by $(-3\cdot144)$ and adapting their result to our
conventions reads\footnote{To check this, one needs to work out the   
supersymmetry variation given in appendix \ref{st} in the gauge (\ref{gfix})   
used in \cite{GrSe} and compare with the corresponding transformations
in appendix A of \cite{GrSe}. In particular, the supersymmetry transformation
rule of the dilaton differs by a factor of $2 i$, and of the vielbein   
by a sign.}
\be \label{3.5}
D_{11} f^{(11,-11)} = 2 i f^{(12,-12)}
\ee
where $D_{11}$ is a modular covariant derivative.   
For later use, we introduce the modular covariant derivatives   
\be
D_{w} = i (\tau_2 {\pa \over \pa \tau} - i {w \over 2}), \qquad
D^*_{\hat{w}} = -i (\tau_2 {\pa \over \pa \tau^*} + i {\hat{w} \over 2}),
\ee
$D_w$ and $D^*_{\hat{w}}$ acting on a modular form  of weight   
$(w,\hat{w})$ (see footnote   
\ref{modular} for the definition) gives a form of weight   
$(w+1,\hat{w}-1)$ and $(w-1,\hat{w}+1)$, respectively.
Our superpotential term is not supersymmetric. We shall show,
however, that our coefficients automatically satisfy 
(\ref{3.5}). In particular, we will show that (\ref{constr}) 
is exactly (\ref{3.5}). Our computations so far were all
done in the gauge invariant formulation,
but the ones in \cite{GrSe} in a specific gauge, so to   
compare our formulas with theirs we first need to express
our results in the gauge used in \cite{GrSe}. 

\subsection{Gauge fixing the $U(1)$}

We discuss in this subsection how to express our results in the
gauge used in \cite{GrSe}. This gauge is   
described in a real basis in section 2.1 of \cite{Gaberdiel:1998ui}.
Expressing (\ref{real}) in $SL(2,R)$ variables using (\ref{sl2b})
and comparing with the results of section 2.1 of \cite{Gaberdiel:1998ui}   
we find that the gauge fixing condition is
\be \label{gfix}
\cos \phi={1 + a_1 \over \sqrt{(1+a_1)^2 + a_2^2}}.
\ee
It will be convenient in what follows to consider $u$ and $v$   
as the independent variables. Then in the gauge (\ref{gfix})
we have
\be \label{gfix2}
u^*={1 \over u-v} - v, \qquad v^* = {1 \over u-v} - u
\ee
We next work out the modular covariant derivatives in terms
of these independent variables,
\bea
D_w &=& -{1 \over 4} \left((2 u - v) {\pa \over \pa v} + u {\pa \over \pa u}
\right) +   
{w \over 2} \nn
D^*_w &=& -{1 \over 4} \left((2 v - u) {\pa \over \pa u} + v {\pa \over \pa v}
\right) +   
{w \over 2}
\eea
Recalling that $F(v)$ is a function of two variables $v$ and
$u^*$ but with $u^*=A v +B$ (see (\ref{fv})), we can further manipulate   
these formulas as
\bea
D_w f(v,u^*=A+Bv) &=& \left[-{1 \over 2} u \pa_v f(v,u^*)   
+ P f(v,u^*) +{w \over 2} f(v,u^*) \right]_{u^*=A + B v} \nn
D^*_w f(v,u^*=A+Bv) &=&   
\left[Q f(v,u^*) +{w \over 2} f(v,u^*) \right]_{u^*=A + B v}
\eea
where   
\bea
P&=&{1 \over 4} \left(v \pa_v - u \pa_u - v^* \pa_v^* \right) \nn
Q&=&-{1 \over 4} \left((2 v -u) \pa_u 
- v \pa_v + (2 u^* -v^*) \pa_v^* \right)   
\eea
where after the differentiation one is instructed to substitute
$u^*=A + B v$ and impose the gauge fixed relations (\ref{gfix2}).
The operators $P$ and $Q$ satisfy the following relations,
\bea \label{PQrel}
P u^* = {1 \over 4} u^*, &&\qquad Q u^* = u^*, \nn
P (u^* v^*) =0, &&\qquad  Q (u^* v^*) = 2 (u^*)^2, \nn
P a = 0 &&\qquad Q a =0.
\eea

\subsection{$SL(2,Z)$ invariance}

After this detour we go back to the computation   
of $t^{(11,-11)}$ and $t^{(12,-12)}$.
Using the definition in (\ref{fv}) we obtain   
\bea \label{t1112}
t^{(11,-11)} &=& - 4 i 2^{16} {1 \over u^*{}^{22}}   
\sum_{n=0}^7 c_n^{(11)} (u^* v^*)^n \tilde{W}^{(15-n)}(a) \nn
t^{(12,-12)} &=& 2^{16} {1 \over u^*{}^{24}}   
\sum_{n=0}^6 c_n^{(12)} (u^* v^*)^n \tilde{W}^{(16-n)}(a)   
\eea
where the combinatorial coefficients are given by
\bea \label{coef11}
&&c_0^{(11)}=1, \quad c_1^{(11)}=-90, \quad c_2^{(11)}=3150,   
\quad c_3^{(11)}=-54600,
\quad c_4^{(11)}=491400, \nn
&&c_5^{(11)}=-2162160, \quad c_6^{(11)}=3603600, \\
&&c_0^{(12)}=1, \quad c_1^{(12)}=-112, \quad c_2^{(12)}=5040,   
\quad c_3^{(12)}=-117600,
\quad c_4^{(12)}=1528800, \nn
&&c_5^{(12)}=-11007360, \quad c_6^{(12)}=40360320,   
\quad c_7^{(12)}=-57657600.    
\eea

Notice that the overall factors of $u^*$ in (\ref{t1112})
carry the local $U(1)$ charge of $t^{(w,-w)}$. Recall   
that $\l$ and $\psi^*$  carry local $U(1)$ charge $3/2$ and   
$-1/2$, respectively. Thus the $\l^{16}$ and $\psi^* \l^{15}$
have $U(1)$ charge $24$ and $22$, respectively. It follows
that $t^{(12,-12)}$ and $t^{(11,-11)}$ should have local   
$U(1)$ charge $-24$ and $-22$, and this is indeed the case.

It is now easy to check that   
(\ref{3.5}) follows from (\ref{constr}). Indeed,
\bea \label{ver35}
D_{11} t^{(11,-11)} &=& -{1 \over 2} u \pa_v t^{(11,-11)}
+ \left(P + {11 \over 2} \right) t^{(11,-11)} \nn
&=& 2 i t^{(12,-12)}
\eea
where we used (\ref{constr}) and the fact that $t^{(w,-w)}$
is an eigenfunction of $P$ with eigenvalue $-w/2$.   
This follows from (\ref{PQrel}), and it is independent   
of the combinatorial factors $c_n^{(w)}$. In this   
respect, gauge invariance is crucial in getting
(\ref{3.5}). Notice also that it was crucial to incorporate the 
contribution of the measure. The numerical coefficient in 
(\ref{constr}) and thus (\ref{ver35}) depends on 
this contribution.

Let us now examine whether the coefficients are eigenfunctions   
of the appropriate Laplacian. Notice that for $(w,-w)$ forms one   
can define two Laplacians,
\bea \label{nablaf}
\nabla_{(-)w}^2 &=& 4 D_{w-1} D_{-w}^* =   
4 \tau_2^2 {\pa \over \pa \tau}  {\pa \over \pa \tau^*}   
- 2 i w \tau_2 ({\pa \over \pa \tau} + {\pa \over \pa \tau^*}) - w (w-1) \nn
\nabla_{(+)w}^2 &=& 4 D^*_{-w-1} D_{w} =   
4 \tau_2^2 {\pa \over \pa \tau}  {\pa \over \pa \tau^*}   
- 2 i w \tau_2 ({\pa \over \pa \tau} + {\pa \over \pa \tau^*}) - w (w+1)   
\eea
and the eigenfunctions satisfy
\bea
\nabla_{(-)w}^2 t^{(w,-w)} &=& \s_w t^{(w,-w)} \nn
\nabla_{(+)w-1}^2 t^{(w-1,-w+1)} &=& \s_w t^{(w-1,-w+1)}.
\eea   
The easiest way to check that (\ref{t1112}) are eigenfunctions
is to compute $D^*_{-12} t^{(12,-12)}$. A short computation using
(\ref{PQrel}) shows that,   
$D^*_{-12} t^{(12,-12)} = C t^{(11,-11)}$,
if and only if the combinatorial factors satisfy
\be
c^{(11)}_n = - {i \over 8 C} (n+1) c^{(12)}_{n+1}   
\ee
Inspection of the coefficients in (\ref{coef11}) shows that this is   
indeed the case with $C=14 i$, and we get
\be \label{d12}
D^*_{-12} t^{(12,-12)} =  14 i t^{(11,-11)}~.
\ee

It follows from these results that   
\bea
\nabla^2_{(-)12} t^{(12,-12)} &=& -112\, t^{(12,-12)} \nn
\nabla^2_{(-)11} t^{(11,-11)} &=& - 90\, t^{(11,-11)}. \label{eigen}
\eea
These results were also checked by a direct computation
using the expression of the Laplacian in (\ref{nablaf}).
Notice that the eigenvalue of the Laplacian in these    
cases, as well as for $t^{(0,0)}$, is given by $c_1^{(w)}$.   
It can be seen from (\ref{naw}) and the fact that the superpotential   
$\tilde{W}$ is a holomorphic function of $a$ that   
$c_1^{(w)}$ has to be equal to the eigenvalue.
Of course, the remaining combinatorial coefficients   
should be consistent with this fact too.

As is discussed in section 2.2 of \cite{GrSe}, the   
eigenvalues of two modular forms that are related to each   
other by modular covariant derivatives are related in a specific
manner. In particular, if $t^{(w-m,-w+m)}$ is related to
$t^{(w,-w)}$ by the application of $m$ modular covariant derivatives
and  $\s_w$ is the eigenvalue of $t^{(w,-w)}$ then
\be \label{rec}
\nabla^2_{(-) w-m} t^{(w-m,-w+m)} = (\s_w + 2 m w - m^2 - m) t^{(w-m,-w+m)}
\ee
The case relevant for us is $w=0$ and $m=-11$ and $m=-12$.
We have computed earlier that $\s_0 = 20$ 
(see (\ref{20})). Applying (\ref{rec}) 
we precisely get (\ref{eigen})!   

To summarize, we have computed the coefficient $t^{(0,0)}$,   
$t^{(11,-11)}$ and $t^{(12,-12)}$ of $r^4, \psi^* \l^{15}$   
and $\l^{16}$ as a function of the superpotential
and its derivatives, and we have shown by independent computations
that all three are eigenfunctions of appropriate $SL(2,R)$ Laplacians,
related to each other by modular covariant derivatives,
and the corresponding eigenvalues are consistent with this   
fact. We have also checked that the supersymmetry constraint (\ref{3.5})
derived earlier in \cite{GrSe} is automatically satisfied
in our case.   

The discussion so far was at the supergravity level.
In string theory, the $SL(2,R)$ symmetry is believed to be   
replaced by a local $SL(2,Z)$ symmetry. The effective action
should now be $SL(2,Z)$ symmetric, and as we next discuss
this implies that   
$t^{(0,0)}$, $t^{(11,-11)}$ and $t^{(12,-12)}$ are uniquely
fixed. In particular, the unique (up to multiplicative
constants) non-holomorphic modular form   
which is an eigenfunction of the $SL(2,Z)$ Laplacian
with eigenvalue $20$ is the Eisenstein series \cite{terras}
\be
E_5 (\tau) = \sum_{\g \in \G_\infty \backslash \G} (\mbox{Im} (\g \tau))^5
= \half (\tau_2)^5 \sum_{(m,n)=1} {1 \over |m \tau + n|^{10}}
\ee
where $(m,n)$ is the greatest common divisor of $m$ and $n$ and  
\be
\G_\infty = \{ \left(
\begin{array}{cc}
\pm 1 & n \\
0 & \pm 1   
\end{array}   
\right)
\in SL(2,Z) = \G \}
\ee
($n$ is any integer).
In other words, $E_5$ is obtained by starting from $\tau_2^5$, which   
manifestly is an eigenfunction of the Laplacian, and taking the 
$SL(2,Z)$ orbit.
Since $\tau_2$ is invariant under the $T$ transformations   
$\tau \to \tau +1$ generating $\G_\infty$, the orbit excludes
these elements. We thus conclude that (up to an overall constant)
\be
t^{(0,0)}(\t,\t^*) = E_5(\t,\t^*)~.
\ee

The asymptotic form of $E_5$ is given by \cite{terras}
\be
E_5 = c_1 \tau_2^5 + c_2 \tau_2^{-4}   
+ \sum_{n \neq 0} a_n \tau_2^{1/2} K_{9/2} (2 \pi |n| \tau_2)   
\exp (2 \pi n \tau_1)
\ee
where $c_1, c_2$ and $a_n$ are (known) constants   
(see \cite{terras} p. 208), and $K_{9/2}$ is a modified 
Bessel function of the third kind.
Notice also that there are no
solutions with $c_1=c_2=0$ (i.e. there are no cusp forms).
The analysis so far was in the Einstein frame. Going over to the   
string frame leads to the result (\ref{s3fin}) given in the introduction.
We discuss the significance of this result in the next section.

\section{Discussion} \label{disc}
\setcounter{equation}{0}

We investigated in this paper the construction of a superinvariant
in type IIB supergravity that contains the well-known $R^4$ term
among its components. We looked for a superinvariant that 
can be constructed as an integral over half of IIB on-shell superspace.
The construction involved the following steps.   
We first showed that the type IIB superspace admits a
chiral superfield whose leading component is the dilaton.
The linearized version of this superfield is well-known \cite{HW},
but the non-linear version has not appeared
before (although its existence was undoubtedly known by the   
experts). We presented an iterative construction of   
all its components, and explicitly discussed the first four.
The iterative formulas originate from the solution of the   
Bianchi identities. The second step involves the construction   
of the appropriate chiral measure. We showed that one can   
systematically investigate the existence of a chiral measure 
by requiring supersymmetry. Supersymmetry leads to 32 
conditions, and we showed that 16 of those can be used to
uniquely determine the components of the chiral measure,
but the other 16 should be satisfied automatically.
It turns out that there is an obstruction at the next-to-leading
order at the non-linear level. The obstruction is the same
as the obstruction to the existence of chiral superfields.

IIB supergravity has an $SL(2,R)$ symmetry,
and this symmetry can be realized linearly by the introduction
of an extra auxiliary scalar and an extra $U(1)$ gauge   
invariance. It is useful to keep the extra gauge invariance   
because it constrains the couplings in the effective action.
We constructed an 8-derivative term by integrating   
an arbitrary holomorphic function of the chiral superfields
over half of superspace. Gauge invariance demands that this
function is a product of a $U(1)$ compensator times   
an arbitrary holomorphic function of $a$ (or equivalently of 
$\tau$). The superpotential term is given in (\ref{supera}).   
     
We further discussed the computation of the superpotential in components.
In particular, we computed the complete dependence on the Riemann
curvature (which actually enters only through the Weyl   
tensor $C$). We find that there are possible terms   
of the form $C^k (DF_5)^l F_5^m$, for appropriate $k,l,m$.
In particular, the index contractions in the $R^4$ terms are   
exactly the ones appearing in the four-point graviton computation
\cite{GW} (i.e. they involve the $t_8$ tensor).   

We also computed the moduli dependent coefficients
of the $R^4$, $\l^{16}$ and $\psi^* \l^{15}$.   
We showed that these coefficients are eigenfunctions   
of (appropriate) $SL(2,R)$ Laplacians with correlated eigenvalues.
Imposing $SL(2,Z)$ invariance uniquely fixes these functions.
In particular, the coefficient of the $R^4$ term   
turns out to be the non-holomorphic Eisenstein series $E_5$.
Using the known asymptotics of $E_5$ we obtained that at   
weak coupling the coefficient of the $R^4$ term, in the string frame,   
consists of two ``perturbative'' terms   
$g_s^{-11/2}$ and $g_s^{7/2}$, see (\ref{s3fin}), plus   
non-perturbative corrections.   
As we discussed in the introduction, this asymptotic behavior
cannot be generated by closed string perturbation theory,
or any known non-perturbative effects. Notice   
that for sufficiently small string coupling constant
the $R^4$ term dominates over the ``leading'' $R$   
term. It thus seems that, if such a (non-supersymmetric) $R^4$ term
is present, then in order to stay within the   
regime of the low energy effective actions, the dilaton
would have to be stabilized at some non-zero value that depends
on the curvature of the background. 

As is well known, the $R^4$ term is generated in string   
perturbation theory at tree and one-loop level \cite{GW}.
One may ask what is the superinvariant that is associated   
with these terms. Such a superinvariant has been constructed
at the linearized level, i.e. in terms of linearized on-shell superfields,
for the $T^2$ reduction of the theory  in \cite{Berkovits:1997pj}.
The superinvariant was given as a sum of two ``superpotential''
terms. The first one involves an integral of an arbitrary function
of a chiral superfield over 16 $\theta$'s, and the second is  
an integral of an arbitrary function of a linear superfield
over a different set of 16 $\theta$'s. In total the   
integration involves 24 different $\theta$'s. 
In eight dimensions it is possible to choose two different sets   
of $\theta$'s and construct two different Lorentz invariant   
measures. In ten dimensions, however, there is only one
Lorentz invariant measure that involves 16 $\theta$'s.
This suggests that the oxidation of the eight dimensional
construction to ten dimensions would involve non-scalar
``superpotentials''. In other words, the integrand should
transform under Lorentz transformations   
such that the action is Lorentz invariant.   
Such construction has been presented in \cite{Howe:1981xy},
but in that paper the measure and the integrand transformed
under an internal symmetry.  In our case we would like them   
to transform under the Lorentz group. 
 
Notice that our construction of the projections of $V$
also leads to an iterative construction of all projections
of all other superfields that appear in the solution of 
the Bianchi identities, since the latter
are related to $V$ by the application of fermionic covariant derivatives.
The term we investigated here
involves an arbitrary scalar function of the chiral superfield.
It would be interesting to study more general constructions that 
may involve other superfields.
 
Our discussion so far was in IIB supergravity.   
The method, however, should be applicable
to all other string theories. In particular, on-shell $N=1$   
supergravity \cite{Nilsson:1981bn} is completely determined
by a scalar superfield $\F$. The coupling of the   
latter to Yang-Mills requires an additional vector superfield.   
The superspace of IIA supergravity has been discussed in   
\cite{Carr:1986tk}. In string theories, one can in principle   
compute the higher derivative corrections by means of   
string amplitude computations and $\s$-model computations.   
In M-theory, however,
the absence of a microscopic formulation makes even more
important the construction of superinvariants.   
Eleven dimensional supergravity is described on-shell
by a single superfield $W_{rtsu}$, totally antisymmetric   
in the flat Lorentz indices, that satisfies the   
constraint, $(\G^{rst} D) W_{rtsu} =0$   
\cite{Cremmer:1980ru,Brink:1980az}. The leading
component of $W_{rtsu}$ is the field strength of the
three form. It will be interesting to investigate   
whether one can use $W_{rtsu}$ in order to construct
a superinvariant associated with the higher derivative
corrections to eleven dimensional supergravity.

\section*{Acknowledgments} We would like to thank M. Green,   
and especially N. Berkovits, P. Howe and P. West for discussions and 
correspondence. We are especially grateful to N. Berkovits and P. Howe 
for informing us about their (unpublished) computation regarding the 
obstruction to the existence of the chiral measure.
SdH and KS would like to thank the Physics Department in Swansea
and the Amsterdam Summer Workshop, AS and KS the Isaac Newton Institute,
and KS the Aspen Center for Physics   
for hospitality during the completion of this work. AS also thanks  
the Spinoza Institute for support. SdH would like to thank the Institute for 
Pure and Applied Mathematics for the organization of the workshop ``Conformal 
Field Theory and Applications" during which part of this work was done.
This material is based upon work supported by the National Science
Foundation under Grant No. PHY-9802484, PHY-9819686 and
DOE grant number E-FG03-91ER40662, Task C.
Any opinions, findings, and conclusions or recommendations expressed in
this material are those of the authors and do not necessarily reflect
the views of the National Science Foundation.

\appendix

\section{Notation and conventions} \label{conv}
\setcounter{equation}{0}

We use the notation and conventions of \cite{HW}, but we denote   
complex conjugation by $*$ (rather than a bar), and the $U(1)$   
charges are normalized as in \cite{S}, i.e. they are half of the   
ones given in \cite{HW}.

The superspace coordinates are   
\be
z^M = (x^m, \th^\mu, \th^{\bar{\mu}})
\ee
where $m= 1,\ldots, 10$ are the 10 bosonic coordinates, and $\theta^{\mu}$
$\mu =1, \ldots, 16$ are 16 complex fermionic coordinates.  They   
form a 16 component Weyl spinor of $SO(9, 1)$. We use the notation
$(\th^{\mu})^* = \th^{\bar{\mu}}$.
Curved space vector and spinor indices are denoted by $m, n,\ldots$ and
$\mu, \nu,\ldots$, respectively, while for tangent space indices we use
$a, b,\ldots$ and $\a, \b,\ldots$

The vielbein one-form superfield is   
\be
E^A = d z^M E_M^A
\ee
with non-zero lowest $\th=0$ components
\bea
E_m^a| &=& e_m^a \nn
E_m^\a| &=& \psi_m^\a  \qquad    
E_m^{\bar{\a}}| = \psi_m^{* \a} \nn
E_{\m}^{\a}| &=& \d_{\m}^{\a} \qquad
E_{\bar{\m}}^{\bar{\a}}| = - \d_{\bar{\m}}^{\bar{\a}}~.
\eea
Here $e_m^a$ is the bosonic vielbein, and $\psi_m^{\a}$ is the gravitino.

A $p$-form may be written as
\bea
&&\f=E^{A_p...A_1} \f_{A_1...A_p} = E^{M_p...M_1} \f_{M_1...M_p} \nn
&&E^{A_p...A_1}=E^{A_p} \wedge \cdots \wedge E^{A_1}, \qquad
E^{M_p...M_1}=dz^{M_p} \wedge \cdots \wedge d z^{M_1}.
\eea
The exterior derivative is given by
\be
d \f =  E^{M_{p+1}...M_1} \pa_{M_1} \f_{M_2 ... M_{p+1}}
=E^{A_{p+1}...A_1}(D_{A_1} \f_{A_2...A_{p+1}}   
+ \half T_{A_1 A_2}{}^B \f_{B A_3 ... A_{p+1}}).      
\ee

The superspace scalar product of two vectors is 
\be
U^A V_A = U^a V_a + U^{\a} V_{\a} - U^{\bar{\a}} V_{\bar{\a}}
\ee
where the minus sign in the last term is for reality of the scalar product.

Following \cite{HW}, we use a mostly minus metric, 
\be
\eta_{ab} = {\rm diag} (+1, -1, \ldots -1)
\ee
and we use the $\G$ matrices 
\bea \label{gamma}
\G^0 &=& \sigma_1 \otimes 1_{16} \nn
\G^i &=& i \sigma_2 \otimes \g^i \quad i= 1, \ldots, 9 \nn
\G_{11} &=&  \sigma_3 \otimes 1_{16}
\eea
where $\sigma_k$ are the Pauli matrices, 
and $\g^i$ are the 16 dimensional hermitian 
gamma matrices matrices.
The 16-dimensional gamma matrices $\g^i$, $i=1,\ldots, 8$,
decompose under SO(8) as  
follows:
\bea \label{so8}
\g^i &=& \left(
\begin{array}{cc}
0 & \g^i_{a \da} \\
\tg^i_{\da a} & 0
\end{array}
\right),
\eea
where $a$ and $\dot{a}$ are {\bf$8_c$} and {\bf$8_s$} 
indices,
respectively. The matrices $\g^i_{a \da}$ are given in 
appendix 5.B of \cite{GSW}. This decomposition 
will be useful in appendix E. The ninth $16\times16$ gamma matrix is given by
\be
\g^9=\g_1\g_2\ldots\g_8=\left(
\begin{array}{cc}
1_8 & 0 \\
0 & -1_8
\end{array}
\right).
\ee
The representation (\ref{gamma}) can be 
rewritten as
\bea   
\G^i &=& \left(
\begin{array}{cc}
0 & \g^i \\
\hat\g^{i} & 0
\end{array}
\right), \quad i= 0, \ldots, 9 \nn
\G_{11} &=&  \sigma_3 \otimes 1_{16}
\le{gammarep}
where $\hat\g$ is defined by (\ref{gamma}). The $\g$'s used in the text 
are the $\g^i$ given here.
The antisymmetrized product of $\g$ matrices has the property that   
$\g_a$, $\g_{abcd}$ and $\g_{abcde}$ are symmetric, and $\g_{ab}$ and
$\g_{abc}$ are antisymmetric. In all explicit formulas it is only
$\g$ that appears.

For arbitrary $32\times32$ matrices $M,N$ and 32-dimensional complex spinors   
$\chi,\l$, the Fierz rearrangement formula reads \cite{Bergshoeff:1981um}:
\be \label{genFie}
M\chi\bar\l N=-{1\over32}\sum_{n=0}^5c_n(\bar\l N\G^{(n)}M\chi)\G^{(n)},
\ee
where $c_0=2, c_1=2, c_2=-1, c_3=-{1\over3}, c_4={1\over12}$ and   
$c_5={1\over120}$ and $\G^{(n)}=\G^{i_1\cdots i_n}$.

One may derive from (\ref{genFie}) useful lemmas. Consider a complex
chiral spinor $\l$. The only non-vanishing bilinear is 
the axial current $\bar\l^*\G^{\m\n\s}\l$.
Combining results obtained by Fierzing $\bar\l^*   
\G^{\m\n\s}\l \bar\l^*$   
and $\bar\l^*\G^{\l[\m\n}\l\bar\l^*\G_\l\G^{\s]}$,
we obtain
\be
\bar\l^*\G^{\m\n\s}\l\bar\l^*=\half\,\bar\l^*\G^{\l[\m\n}\l\bar\l^*   
\G_\l\G^{\s]}~.
\ee
This is the analog of (A.3) in \cite{Bergshoeff:1981um}
but for complex spinors. Now multiplying from the right by
$\G^{(3)}\l$ we get
\bea
\bar\l^*\G^{abc}\l\bar\l^*\G^d{}_{bc}\l=0~.
\le{Fierzing}
Using this formula we will show in appendix E that 
the fourth projection of the $V$ contains only the Weyl tensor.
 
For the graded commutator of two covariant derivatives we have
\be
[D_A, D_B \} = - T_{A B}{}^C D_C   
+ \frac{1}{2} R_{ABC}{}^D L_D{}^C + 2i M_{AB} \k~,
\ee
where $T_{AB}{}^C$ is the torsion, $R_{ABC}{}^D$ is the curvature tensor,
$M_{AB}$ is the $U(1)$ curvature, $\k$ is the $U(1)$ generator, and   
$L_{AB}$ are the generators of $SO(9,1)$
\be
L_{ab} = -L_{ba}   
\qquad L_\a{}^\b =\frac{1}{4} (\g^{ab}{})_{\a}{}^{\b} L_{ab},
\qquad L_{\bar{\a}}{}^{\bar{\b}} = - L_{\a}{}^{ \b}, \qquad
L_{\a b}=L_{a \b} = L_\a{}^{\bar{\b}}=L_{\bar{\a}}{}^\b=0.
\ee
The covariant derivative  acting on a field $\phi_B$ of $U(1)$ charge
$q$ is given by
\bea
D_A \f_B &=& E_A^M D_M \f_B
= E_A^M (\partial_M \phi_B + \O_{MB}{}^C \f_C + 2 i q Q_M \f_B) \nn
&=&\partial_A \phi_B + \O_{AB}{}^C \f_C + 2 i q Q_A \f_B~.
\eea
The spin-connection $\O_B{}^C = dz^M \O_{MA}{}^B$ has the same symmetry   
properties as $L_A{}^B$.
   
Complex conjugation (more properly Hermitian conjugation)   
reverses the order of elements,
\be
(A B)^* = B^* A^*.
\ee
It acts on derivatives as
\be
(\partial_A)^* = - (-)^A \pa_A.
\ee

\section{IIB supergeometry} \label{bianchi}
\setcounter{equation}{0}

In \cite{HW} the torsion and curvature tensors are
computed from Bianchi identities. We list here without derivation
all non-zero components. We closely follow their notations.
\bea
T_{\a \bar{\b}}^c &=& -i \g_{\a \bar{\b}}^c \nn
T_{\a \b}^{\bar{\g}} &=& (\g^{a})_{\a \b} (\g_a)^{\g \d} \L^*_\d -   
2 \d_{(\a}^{\g} \L^*_{\b)} \nn
T_{a \b}^{\bar{\g}} &=& - \frac{3}{16} (\g^{bc})_{\b}{}^{\g} F^*_{abc}
- \frac{1}{48} (\g_{abcd})_{\b}{}^{\g} F^{*bcd}  \nn
T_{a \b}^{\g} &=& i \left( \frac{21}{2} X_a \d_{\b}^{\g} +   
\frac{3}{2} (\g_{ab})_{\b}{}^{\g} X^b   
+ \frac{5}{4} (\g^{bc})_{\b}{}^{\g} X_{abc}   
+ \frac{1}{4} (\g_{abcd})_{\b}{}^{\g} X^{bcd} +   
(\g^{bcde})_{\b}{}^{\g} Z_{abcde} \right) \nn
T_{ab}^{\a} &=& \Psi_{ab}^{\a} \nn
T_{a \bar{\b}}^{\bar{\g}} &=& - (T_{a \b}{}^{\g})^* \nn
T_{a \bar{\b}}^{\g} &=& - (T_{a \b}{}^{\bar{\g}})^*~,    
\eea
with
\bea
Z_{abcde} &=& Z^+_{abcde} + {1 \over 48} X_{abcde}=
\frac{1}{192} F_{abcde} + \frac{1}{16} X_{abcde} \nn
X_{a_1 ..a_i} &=& \frac{1}{16} \L^* \g_{a_1 ..a_i} \L \label{Z}~.
\eea

The nonzero curvature components are
\bea
R_{\a \b ab} &=& i \left( \frac{3}{4} (\g^c)_{\a \b} F^*_{abc}   
+ \frac{1}{24} (\g_{abcde})_{\a \b} F^{*cde} \right) \nn
R_{\a \bar{\b} a b} &=& - \left( 3 (\g_{abc})_{\a \b} X^c + 5 (\g^c)_{\a \b}
X_{abc} + \frac{1}{2} (\g_{abcde})_{\a \b} X^{cde} \right. \nn
&&+ \left. \frac{1}{2} (\g^{cde})_{\a \b}   
\left(\frac{1}{12} F_{abcde} + X_{abcde} \right) \right) \nn
R_{\a \bar{\b} \g \d} &=& \frac{1}{4} (\g^{ab})_{\g \d} R_{\a \bar{\b}   
a b} \nn
R_{\a \b \g \d} &=& \frac{1}{4} (\g^{ab})_{\g \d} R_{\a \b   
a b} \nn
R_{\a b cd} &=& - \frac{1}{2} i \left( (\g_b)_{\a \b} \Psi_{cd}{}^{\bar{\b}}
+ (\g_c)_{\a \b} \Psi_{bd}{}^{\bar{\b}} - (\g_d)_{\a \b} \Psi_{bc}{}^{\bar{\b}}
\right),
\eea
and the non-zero U(1) curvature components are
\bea
M_{\a \bar{\b}} &=& 2 i \L_{\a} \L^*_{\b } \nn
M_{\a b} &=& - i \bar{P}_{b} \L_{\a} \nn
M_{ab} &=& -i \bar{P}_{[a} P_{b]}~.
\eea

The non-zero components of the tensors $P_A$, $F_{ABC}$ and $F_{ABCDE}$ are
\bea
P_{\a} &=& - 2 \L_{\a}, \qquad P_a \nn
F_{a \b \g} &=& F^*_{a \bar{\b} \bar{\g}} = -i (\g_a)_{\b \g}, 
\qquad F_{abc} \nn   
F_{ab \bar{\g}} &=& - (\g_{ab})_{\g}{}^{\d} \L_{\d} \qquad F^*_{ab \g} 
= -(\g_{ab})_{\g}{}^{\d}
\L^*_{\d} \nn
F_{abc \a \bar{\b}} &=& (\g_{abc})_{\a \b} \nn
F_{abcde} &=& F^+_{abcde} - 8 X_{abcde}\, ,
\eea
where $F^+$ denotes the self-dual part.

\section{Supersymmetry transformation rules} \label{st}
\setcounter{equation}{0}

In this appendix we list the supersymmetry transformation rules. 
In these formulae, we corrected several typos in the original paper of Howe
and West \cite{HW}\footnote{We thank Paul Howe and Peter West for confirming
the corrections.}, but we note that despite considerable effort
we were unable to exactly match the transformation rules below to the ones
reported in \cite{SW,S}, indicating that there maybe typos in 
\cite{SW,S} and/or in the solution of Bianchi's given in the 
previous appendix.  
\be 
\d e_{m}{}^a = - i \left( (\zeta^* \g^a \psi_m ) +   
(\zeta \g^a \psi^*_m ) \right)
\ee
\bea
\d \psi_m &=& \nabla_m \zeta - \frac{3}{16} \hat{f}_{mab} \g^{ab} \zeta^*   
+ \frac{1}{48} \hat{f}^{bcd} \g_{mbcd} \zeta^*   
- \frac{1}{192} i \hat{g}_{mabcd} \g^{abcd} \zeta \nn
&&+ \frac{1}{16} i \left[ - \frac{21}{2} (\l^* \g_m \l)   
+ \frac{3}{2} (\l^* \g^a \l) \g_{ma} +   
\frac{5}{4} ( \l^* \g_{mab} \l) \g^{ab}   
- \frac{1}{4} (\l^* \g^{abc} \l) \g_{mabc} \right. \nn
&&\left. - \frac{1}{16} (\l^* \g_{mabcd} \l) \g^{abcd} \right] \zeta   
- (\zeta^* \g^a \psi^*) (\g_a \l) + (\psi^*_m \l) \zeta^*   
- (\zeta^* \l) \psi^*_m
\eea
\be   
\d u = 2 (\zeta^* \l^*) v \qquad \d v = - 2 (\zeta \l) u   
\ee
\be
\d \l = \frac{1}{24} i \hat{f}_{abc} \g^{abc} \zeta + \frac{1}{2} i
\hat{p}_a \g^a \zeta^*
\ee
\be
\d (a^*_{mn}, a_{mn}) = - \left( (\zeta \g_{mn} \l^*)  +   
2 i (\zeta^* \g_{[m} \psi^*_{n]} ) , - (\zeta^* \g_{mn} \l) +
2 i (\zeta \g_{[m} \psi_{n]}) \right)  \cv^{-1}
\ee
\be
\d b_{mnrs} = - 4 (\zeta \g_{[mnr} \psi^*_{s]} ) + 
4 (\zeta^* \g_{[mnr} \psi_{s]} ) + 
12 i \left( a_{[mn} \d a^*_{rs]} - a^*_{[mn} \d a_{rs]} \right)
\ee
where $\cv$ is given in (\ref{nu}).

Here $\hat{p}_a$, $\hat{f}_{abc}$ and $\hat{f}_{abcde}$ are the leading
components of the corresponding superfields
\bea
P_a| &=& \hat{p}_a = p_a + 2 (\psi_a \l) \nn
F_{abc}| &=& \hat{f}_{abc} = f_{abc} - 3 (\psi^*_{[a} \g_{bc]} \l)   
- 3 i (\psi_{[a} \g_b \psi_{c]}) \nn
F_{abcde}| &=& \hat{f}_{abcde} = f_{abcde} + 20   
(\psi^*_{[a} \g_{bcd} \psi_{e]})~.
\eea

\section{$F_5^2$ terms in the dilaton superfield} \label{NL}
\setcounter{equation}{0}

We provide in this appendix some details leading to (\ref{f5nl}). 
Starting from (\ref{d4v}) and collecting only the $F_5 F_5$ terms we get
\bea
D_{[\d} D_{\g} D_{\b} D_{\a]} V|_{f_5 f_5}     
&=& -\frac{1}{4} u \left\{
 \frac{1}{6912} (\g^a{}_{bc})_{[\b \a} (\g^{ijk})_{\g \d]}
\e_{aijkmnpm'n'p'} f^{bmnpl} f^{cm'n'p'}{}_l \right. \nn
&-& \left. \frac{1}{192} 
(\g^{abc})_{[\b \a} (\g_a{}^{de})_{\g \d]}
f_{bemnp} f_{cd}{}^{mnp} \right\}. \label{9}
\eea
To derive this we expressed products of gamma matrices into 
completely antisymmetric combinations and we used the fact that 
only $\g^{ab}$ and $\g^{abc}$ (and $\g^{(8)}$ and $\g^{(7)}$)
are antisymmetric in the spinor indices. Furthermore, we used the gamma 
matrix identity 
$\g_{a_1..a_7} = -{1 \over 3!} \e_{a_1 .. a_7 b_1 b_2 b_3} \g^{b_1 b_2 b_3}$
in the first term on the right hand.

Let us define
\bea
A &=& (\g^{abc})_{[\b \a} (\g^{def})_{\g \d]} f_{bafmn}
f_{ced}{}^{mn}  \nn
B &=&  (\g^{abc})_{[\b \a} (\g_a{}^{de})_{\g \d]}
f_{bemnp} f_{cd}{}^{mnp} \nn
C&=&(\g^{abc})_{[\b \a} (\g_{a}{}_{de})_{\g \d]} f_{bcmnp} f^{demnp}\nn
D&=&(\g^{abc})_{[\b \a} (\g_{def})_{\g \d]} f_{abcmn} f^{defmn} \nn
E &=& (\g^a{}_{bc})_{[\b \a} (\g^{ijk})_{\g \d]}
\e_{aijkmnpm'n'p'} f^{bmnpl} f^{cm'n'p'}{}_l~. 
\eea
$B$ and $E$  appear in (\ref{9}).

$B$ and $C$ can be shown to be equal to zero due to the 
self-duality of $f_{(5)}$.
We show this for $C$,
\bea
C &=& \g^{abc} \g_{a}{}_{de} f_{bcmnp} f^{demnp} \nn
&=&    
\g^{abc} \g_{ade} \frac{1}{(5!)^2} \e_{bcmnpa_1..a_5} \e^{demnpb_1..b_5}
f^{a1..a5} f_{b1..b5} \nonu
&=& -\g^{abc} \g_{a}{}_{de}  \frac{3!}{(5!)^2} \d_{bca_1..a_5}^{deb_1..b_5}
f^{a1..a5} f_{b1..b5} \nonu
&=& - \g^{abc} \g_{a}{}_{de} \frac{2! 3!}{5!} \left( \d_{bc}^{de}    
\d_{a_1..a_5}^{b_1..b_5}  - 10 \d_{bc}^{d b_1} \d_{a_1...a_5}^{e b_2..b_5}
+ 10 \d_{bc}^{b_1 b_2} \d_{a_1...a_5}^{de b_3..b_5} \right)    
f^{a1..a5} f_{b1..b5} \nonu
&=& - \g^{abc} \g_{a}{}_{de} f_{bcmnp} f^{demnp} \nn
&=& - C~,
\eea
where we supress the spinor indices. Notice that we used (\ref{Fierzing}),
i.e. $(\g^{abc})_{[\b \a} (\g_{abd})_{\g \d]}=0$. A similar computation
establishes that $B=0$. Using the self-duality of $f_{(5)}$ in $A$ and $D$ one 
finds again $B=C=0$ (and no constraint on $A$ and $D$).
Furthermore, similar manipulations yield
\bea
E &=& 18 (3A+2B+C-D).
\eea
We thus obtain
\be
E = 18 \g^{abc} \g^{def} (3 f_{bafmn} f_{ced}{}^{mn}  
- f_{abcmn} f_{def}{}^{mn})~.  
\ee
Combining these formulas we finally get 
\be 
D_{[\d} D_{\g} D_{\b} D_{\a]} V|_{f_5 f_5} =    
-\frac{u}{1536}  
\g^{abc}_{[\b \a} \g^{def}_{\g \d]}
(3 f_{bafmn} f_{ced}{}^{mn}  
- f_{abcmn} f_{def}{}^{mn})
\ee
which is (\ref{f5nl}).

\section{$R^4$ term in the superpotential} \label{r4}
\setcounter{equation}{0}

In this appendix we show that the $R^4$ term appearing
among the components
of the superpotential term has the same structure as the usual
$R^4$ term that appears in string scattering amplitudes. 
The computation presented here follows closely the analysis 
in \cite{Green} and it is included for completeness.

In section 6 
we showed that this term comes from the fourth projection of the scalar 
superfield. The relevant terms are given by
\be
S_{r^4} =  \int d^{10} x\, e {1 \over (4!)^5}
\e^{\a_1...\a_{16}} (D_{\a_1} .. D_{\a_4} V|)  
(D_{\a_{5}} .. D_{\a_{8}} V|) (D_{\a_{9}} .. D_{\a_{12}} V|)
(D_{\a_{13}} .. D_{\a_{16}} V|) 
\ee
Using (\ref{car}) and keeping only the term proportional to $c^4$ 
we get
\bea 
S_{r^4}&=&\int d^{10} x e {1 \over (4!)^5 16^4} u^4 F^{(4)}(v)   
\e^{\a_1...\a_{16}}   
(\g^{a i_1 i_2}{}_{\a_1 \a_2}\g_{a}{}^{i_3 i_4}{}_{\a_3 \a_4} 
c_{i_1 i_2 i_3 i_4}) \times \nn
&& \qquad \cdots
(\g^{d l_1 l_2}{}_{\a_{13} \a_{14}}
\g_{d}{}^{ l_3 l_4}{}_{\a_{15} \a_{16}} c_{l_1 l_2 l_3 l_4})
\le{SR4}
One may similarly manipulate the $F_5$-dependent terms, 
but we shall not do this here.

The aim of this appendix is to eliminate the gamma matrices from (\ref{SR4}).
The resulting expression will have only vector indices contracted
among themselves. This can be done in several ways, and we will  
choose to go via a route motivated by the light-cone computations.
A covariant computation is done in appendix B of \cite{Peeters:2001qj}.
We first go to light-cone coordinates by decomposing space-time indices into  
longitudinal and transverse indices, $(+,-,i)$, where $i=1,\ldots,8$. We will  
also decompose SO(9,1) spinors into SO(8) spinors in the {\bf$8_c$} and  
{\bf$8_s$} representations, which we will denote by undotted/dotted indices,  
respectively. So, in the representation \eq{gammarep} a spinor of negative  
chirality, $\left(\begin{array}{cc}0\\1\\ \end{array}\right)\otimes\th$,  
decomposes under SO(8) as $\th=(\th_a,\dot\th_{\dot a})$. The epsilon  
tensor in (\ref{SR4})
also factorizes into a product of two epsilon tensors. Since chiral  
spinors have $8+8$ independent components, we can use them to represent  
the epsilon tensors:
\be
\int d^8\th\,\th^{a_1}\cdots\th^{a_8}=\e^{a_1\cdots a_8},  
\qquad \int d^8\dot\th\,\dot\th^{\dot a_1}\cdots\dot\th^{\dot a_8}  
=\e^{\dot a_1\cdots \dot a_8}~.
\label{integrals}\ee
This will be a useful bookkeeping device that simplifies our intermediate  
expressions. We also decompose the gamma matrices as in (\ref{so8}).
In the remaining of this appendix we deal with  
the $8\times8$-dimensional gamma matrices only.

The expression \eq{SR4} then reduces to
\be
S_{r^4}=\int d^{10}xd^8\th d^8\dot\th\,e\,
{1\over(4!)^5 16^4}\,u^4F^{(4)}(v)\Phi^4[c]
\label{R4}\ee
where
\be
\Phi[c]=2(-\th\g^{ij}\th\dot\th\g^{kl}\dot\th  
+\th\g^{mij}\dot\th\dot\th\g^{mkl}\th)c_{ikjl}~.
\label{Phi}
\ee
This is the transverse contribution, and we have omitted longitudinal  
terms which contribute to covariantize the final expression. This is the  
first line in formula (56) of \cite{Green}, but already rewritten in terms  
of the Weyl tensor. Indeed, either from the SO(9,1) Fierzing or directly in  
light-cone gauge one can check that all the terms that include the Ricci  
tensor cancel out.

To remove the space-time indices $m$ that are summed over in the above  
expression, we perform SO(8) Fierzings:
\be
\th_a \th_b = {1 \over 16} (\th \g_{ij} \th) \g^{ij}_{ab}, \qquad
\dot\th_{\da} \dot\th_{\db}
= {1 \over 16} (\dot\th \tg_{ij} \dot\th) \tg^{ij}_{\da \db}
\ee
and after computing the traces over products of SO(8) $\g$'s we get:
\bea
(\th \g^{ijk} \dot\th) (\dot\th \tg^{lm}{}_k \th) c_{ijlm}
&=&-2(\th\g^{ij}\th)(\dot\th\ti\g^{kl}\dot\th)\, c_{ijkl}~,
\le{fierzing}
and again we suppress terms proportional to traces of the Weyl tensor.  
Since these are traces over transverse indices, they are not zero but rather  
recombine with remaining terms from the longitudinal part to give a vanishing  
contribution to the final result. Therefore the contribution in \eq{fierzing}  
is the one that covariantizes to the final 10d expression. Filling  
\eq{fierzing} into \eq{Phi}, we get
\be
\Phi[c]=-6(\th\g^{ij}\th)(\dot\th\ti\g^{kl}\dot\th) c_{ijkl},
\label{c}
\ee
which is the second line of (56) in \cite{Green} (but
with the overall coefficient corrected).

Filling \eq{c} in the expression for the action \eq{R4}, and using the  
representation \eq{integrals}, we get
\bea
S_{R^4} &=&\int d^{10}x\,e\, t^{(0,0)}\, \e^{a_1 \ldots a_8}  
\e^{\dot a_1\ldots\dot a_8} \,
(\g^{i_1 j_1}_{a_1 a_2} \ti\g^{k_1 l_1}_{\dot a_1 \dot a_2}  
c_{i_1 j_1 k_1 l_1})\cdots
(\g^{i_4 j_4}_{a_7 a_8} \ti\g^{k_4 l_4}_{\dot a_7 \dot a_8}  
c_{i_4 j_4 k_4 l_4})
\eea
where $t^{(0,0)}$ is given in \eq{t00}.

As shown in \cite{Green}, this is equal to
\bea
S_{r^4} &=&\int d^{10}x\,e\, t^{(0,0)} (t_8^{i_1 j_1 \ldots i_4 j_4} +  
\frac{1}{2}   
\e^{i_1 j_1 \ldots i_4 j_4} ) (t_8^{k_1 l_1 \ldots k_4 l_4} -
\frac{1}{2} \e^{k_1 l_1 \ldots k_4 l_4} )   
c_{i_1 j_1 k_1 l_1}\cdots c_{i_4 j_4 k_4 l_4} \nn
&=& \int d^{10}x\,e\,  
t^{(0,0)} \left( t_8^{i_1 j_1 \ldots i_4 j_4} t_8^{k_1 l_1 \ldots k_4 l_4}
- \frac{1}{4} \e^{i_1 j_1 \ldots i_4 j_4}  \e^{k_1 l_1 \ldots k_4 l_4} \right)
c_{i_1 j_1 k_1 l_1}\cdots c_{i_4 j_4 k_4 l_4}
\eea
The tensor $t_8$ is the well-known kinematical factor appearing in the  
tree-level and one-loop string scattering amplitudes. The explicit  
expression can be found in, e.g., appendix 9.A of \cite{GSW}. It is now  
straightforward to write the 10d covariant expression. As advertised, we  
get the usual combination of Weyl tensors:
\be
S_{r^4} = \int d^{10}x\, e\,
t^{(0,0)} \left( c^{hmnk} c_{pmnq} c_h{}^{rsp} c^q{}_{rsk}
+ \frac{1}{2} c^{hkmn} c_{pqmn} c_h{}^{rsp} c^q{}_{rsk} \right).
\ee

\end{document}